\documentclass[pre,twocolumn]{revtex4-1}
\usepackage{graphicx}
\usepackage{tikz}
\usepackage{mathtools}
\usepackage{amssymb}
\usepackage{chngcntr}

\usepackage{tikz}
\usetikzlibrary{calc}
\usetikzlibrary{backgrounds,arrows,decorations.markings,decorations.pathmorphing,positioning}

\usepackage{environ}
\makeatletter
\newsavebox{\measure@tikzpicture}
\NewEnviron{scaletikzpicturetowidth}[1]{%
  \def\tikz@width{#1}%
  \def\tikzscale{1}\begin{lrbox}{\measure@tikzpicture}%
  \BODY
  \end{lrbox}%
  \pgfmathparse{#1/\wd\measure@tikzpicture}%
  \edef\tikzscale{\pgfmathresult}%
  \BODY
}
\makeatother

\tikzstyle{node} = [draw, circle,fill=blue!20, node distance=3cm,
    minimum height=0em]
\tikzstyle{connection}=[inner sep=0,outer sep=0]    
\tikzset{snake it/.style={decorate, decoration=snake}}

\newcommand*\diff{\mathop{}\!\mathrm{d}}

\begin{document}
\title{Monte Carlo Simulation of Long Hard-Sphere Polymer Chains in Two to Five Dimensions}

\author{Stefan Schnabel}

\author{Wolfhard Janke}

\affiliation{Institut f\"ur Theoretische Physik, Universit\"at Leipzig, IPF 231101, 04081 Leipzig, Germany}
\date{\today}
\begin{abstract}

We perform simulations for long hard-sphere polymer chains using a recently developed binary-tree based Monte Carlo method. Systems in two to five dimensions with free and periodic boundary conditions and up to $10^7$ repeat units are considered. We focus on the scaling properties of the end-to-end distance and on the entropy and their dependence on the sphere diameter. To this end new methods for measuring entropy and its derivatives are introduced. By determining the Flory exponent $\nu$ and the weakly universal amplitude ratio of end-to-end distance to radius of gyration we find that the system generally reproduces the behavior of self-avoiding lattice walks in strong support of universality.

\end{abstract}

\maketitle

\section{Introduction} 
\label{sec-intro}

Research into the formation of nanopatterns of macromolecules is key
to an understanding of many of their properties.
This comprises single as well as 
entire assemblies of linear polymers whose properties may range from 
flexible to semiflexible to rather stiff, providing internal constraints. 
The polymers can be of synthetic or biological origin which often
exhibit similar behaviors. In both applications, they
may be described by generic coarse-grained bead-stick or 
bead-spring models with Lennard-Jones-type interactions including 
excluded-volume repulsion among the monomers or beads, but often also 
chemically realistic all-atom models are considered. The characteristic polymer 
conformations governing the pattern formation process are studied in bulk, 
in the presence of structuring surfaces or under confinement conditions, 
acting as external constraints for the structure formation. In numerical
studies, depending on the problem at hand, usually Monte Carlo (MC) or
Molecular Dynamics (MD) computer simulations are employed.

One important focus of our own contributions to this field was on
nonequilibrium pattern formation of polymers that are suddenly
quenched from random-coil to globule conditions. Qualitatively
this involves distinguishing the 
``sausage'' \cite{deGennes_sausage}
and
``pearl-necklace'' \cite{Halperin-Goldbart_pearls}
pictures of the coarsening process. More quantitatively we were
interested in characterising the kinetic scaling laws associated
with this process and related aging properties. Our approach 
strongly relies on an analogy with coarsening and aging of particle 
and spin models \cite{our_pre2019,our_pre2021,our_prl2020} and is 
reviewed in Ref.\ \cite{our_epjb2020a}. For generic models (in implicit 
solvent) we consistently find in our MC simulations that the 
``pearl-necklace'' picture is clearly favored.
This mechanism was also observed to govern
the pathway of the collapse of the polypetide backbone of a protein
in atomistic all-atom MD simulations of polyglycine (in explicit water) 
\cite{our_macromolecules2019}. As preparation for including hydrodynamic 
effects we conducted a study of dissipative dynamics of a single polymer 
in solution using the Lowe-Andersen approach, belonging to the class of 
dissipative particle dynamics (DPD) simulations \cite{our_jpcs2019a}. While 
for low viscosity the ``pearl-necklace'' scenario prevails we do observe
a crossover to the ``sausage'' picture for sufficiently high 
viscosities \cite{our_viscous-tobe1}.

In a related line of research we investigated 
flexible polymers that are endowed with some kind of activity
\cite{our_smat2021,our_sm2022a,our_sm2022b,our_jpcs2022a}.
Here we employed Langevin simulations.
The added activity introduces intrinsic nonequilibrium effects in the collapse 
kinetics at the coil-globule transition and governs the steady-state 
properties of the emerging globular state.

A particularly interesting motif of semiflexible polymer conformations are 
knots of various types that appear to characterize (like an order parameter) 
stable phases of semiflexible polymers. Extending our previous work, we 
studied this peculiar property systematically by comparing bead-stick and 
bead-spring polymer models in dependence on the polymer stiffness and the 
ratio of the (average) bond length to the distance of the Lennard-Jones 
potential minimum and thereby identified favorable conditions for the 
formation of stable knots \cite{ours_macromolecules2021}.

Methodologically, we generalized the relatively new population annealing method
for MC \cite{pa_iba,pa_hukushima-iba,pa_machta,martin2021} to population 
annealing molecular dynamics (PAMD) 
simulations \cite{our_prl2019,our_jpcs2019b,our_jpcs2022b}. For protein studies
this new method may turn out to be superior to the currently usually employed 
parallel tempering simulations, in particular when implemented on massively 
parallel computer architectures (such as graphics processing units (GPUs) 
\cite{lev2017,lev2019}). A benchmark comparison for the B1 domain of protein G
is in preparation \cite{our_proteinG-tobe}.
We also introduced a nonflat histogram technique 
\cite{our_pre2020} that generalizes the commonly employed flat multicanonical
method \cite{muca1,muca2,flat-review_wj-paul} and later adapted this idea 
also to Wang-Landau simulations \cite{wl,flat-review_wj-paul} with nonflat 
distributions \cite{our_cpc2021}. While in Ref.\ \cite{our_pre2020} we 
first drew connection to recent work \cite{our_epjb2020b} and
exemplified the method in a ground-state study for the Edwards-Anderson 
spin-glass model, in the latter paper \cite{our_cpc2021} we explicitly 
demonstrated the usefulness
of the proposed method for unraveling the intriguing low-temperature 
``crystal-like'' patterns of Lennard-Jones polymers. Currently we are employing 
this method for determining the ground-state patterns of lattice peptides
described by the HP model \cite{our_HP-tobe}.

Most relevant for the polymer study presented here is the recent proposal of a
MC algorithm that employs tree-like data structures of the polymer's 
``internal'' degrees of freedom in combination with a ``parsimonious'' 
Metropolis acceptance
criterion \cite{our_cpc2020}. 
While preserving the standard Metropolis dynamics, this algorithm speeds up 
simulations with power-law long-range interactions significantly and hence 
allows the study of much longer macromolecules than before. Subsequently, the 
general setup of the method inspired another ``external'' variant that can 
deal very efficiently with algebraically decaying long-range interactions of
particle and spin systems \cite{our_arxiv2022}. A first concrete application
to the coarsening kinetics of the conserved Ising model has just 
appeared \cite{our_prl2022}.

Our new method for polymers \cite{our_cpc2020} allows efficient 
simulations of the collapse transition of polymers with untruncated Lennard-Jones 
interactions \cite{our_jpcp2022}. As will be discussed below, it
also gives completely new possibilities for investigating seemingly simple 
(athermal) hard-sphere polymers and enables novel observations which would 
have been hardly possible with standard simulation techniques.


An even simpler fundamental model in statistical physics is the self-avoiding lattice walk -- a sequence of steps on a regular lattice that is unable to visit any lattice site more than once. Not only does this model serve as the most basic approximation for polymers with repulsive interaction, it is also a realization of the $O(n)$ vector spin model with $n=0$ and, therefore, its properties are of interest also in higher dimensions. Self-avoiding walks have been studied extensively both analytically and numerically \cite{TSAW,Sokal_book} and are generally well understood. Yet, research is still ongoing and a powerful new MC method has recently been introduced \cite{Clisby2}. Through its application the scaling exponent $\nu$ in three dimensions, which is not known analytically, could be determined up to six digits \cite{Clisby3} significantly improving earlier estimates. For a recent high-precision numerical study of four-dimensional self-avoiding walks see Ref. \cite{Deng_saw_4d}.

The hard-sphere polymer considered here is the simplest generalization of a self-avoiding walk to an off-lattice geometry. Although it can be understood as a walk, a sequence of steps of a certain length in random directions where any new position must be a distance $d$ away from previous positions, the more common picture is that of a linear chain of non-overlapping hard spheres of diameter $d$ that are connected by bonds of fixed length. The universal aspects of the behavior of self-avoiding walks, e.g., scaling exponents, should, of course, apply to this model, too. Particularities that arise from the lattice geometry on the other hand are absent. Another important difference is that for the hard-sphere polymer the strength of the repulsion can be influenced directly by the choice of the sphere diameter.
For the lattice walk this is only possible indirectly for instance by selecting a particular lattice type or by allowing the walk to directly jump to next-nearest neighbors of the currently occupied site as well.

Hard-sphere polymers have been investigated by means of MC simulations some time ago \cite{Kremer}, but due to limits in hardware and methods only chains of length $60$ could be investigated. In an earlier publication \cite{our_cpc2020} we adapted the aforementioned new lattice algorithm to continuous degrees of freedom and are now able to simulate polymers with millions of repeat units. This qualitative difference in the size of the system leads to much smaller corrections to scaling and the predicted behavior manifests itself more clearly.

In this study we simulate hard-sphere polymers in two to five dimensions and compare the results with theory. We vary the sphere diameter and test predictions for crossover scaling and observe the resulting change in entropy.

The rest of the paper is organized as follows: In section \ref{sec-ModObs} we first briefly define the model, discuss the observables we are interested in, and then introduce the concept of periodic boundary conditions for off-lattice polymers. This is followed in section \ref{sec-Meth} by a short explanation of the MC methods we use in this study. The first is the binary tree method \cite{Clisby2} for the efficient implementation of the pivot algorithm \cite{Sokal} adapted to hard-sphere polymers and the second is a novel specialized flat-histogram method that allows the measurement of entropy. The main part of the paper is section \ref{sec-Res} where we present the results. We first discuss geometric quantities like the end-to-end distance and the radius of gyration in two to five dimensions and then analyze the entropy and its dependence on dimensionality and sphere diameter. We close the paper with some conclusions in section \ref{sec-Conc}.


\section{Model and observables}
\label{sec-ModObs}

\subsection{Hard-sphere polymer}
The model we consider in this study is the off-lattice version of a self-avoiding walk (SAW). It is a fully flexible chain with $N$ monomers at positions $\mathbf{X}=(\mathbf{x}_1,\dots,\mathbf{x}_N)$ in $D$ dimensions where the monomers are connected by rod-like bonds of fixed length:
\begin{equation}
|\mathbf{x}_k-\mathbf{x}_{k-1}| = b.
\end{equation}
In the following we set $b=1$, which is equivalent to expressing all distances in units of $b$.

The monomers themselves are hard spheres with a diameter $d\le 1$:
\begin{equation}
|\mathbf{x}_i-\mathbf{x}_j| \ge d \quad \text{for all} \quad i\ne j.
\label{eqn:excl_vol}
\end{equation}

\subsection{Observables}

Two geometric observables are commonly considered: The end-to-end distance $R=|\mathbf{x}_N-\mathbf{x}_1|$ and the squared radius of gyration
\begin{equation}
R_{\rm gyr}^2=\frac1N\sum\limits_{i=1}^N\left(\mathbf{x}_i - \frac1N\sum\limits_{j=1}^N\mathbf{x}_j\right)^2.
\end{equation}
Note that in this study we refer with $N$ to the number of monomers or beads while the number of bonds (or steps in the context of SAWs) is labeled as $L=N-1$.

Another important property of SAWs is the number of possible walks $c_L$ with $L$ steps (bonds).
On a lattice for finite $L$, $c_L$ is an integer number that can at least in principle easily be determined simply by counting all walks. In the off-lattice case the situation is more complicated since the equivalent quantity is the integral over the accessible state space which we denote $e ^S$ implicitly defining the chain's entropy $S$. Formally it is:
\begin{equation}
e^S=\int\prod\limits_{i=2}^{N}\diff \mathbf x_i \prod\limits_{i=2}^{N}\delta(|\mathbf x_i - \mathbf x_{i-1}|-b)\prod\limits_{i=1}^{N-2}\prod\limits_{j=i+2}^{N}\Theta(|\mathbf{x}_i-\mathbf{x}_{j}|-d).
\label{eq:S_def}
\end{equation}
The restrictions of fixed bond length and excluded volume are represented by delta-distributions $\delta$ and Heaviside functions $\Theta$, respectively, and the position of the first monomer $\mathbf x_1$ is arbitrary due to the translational symmetry of the polymer as a whole.
For the pure random walk ($d=0$) it is $e^S={\cal S}_D^{N-1}={\cal S}_D^L$ with the surface area ${\cal S}_D$ of the $D$-dimensional sphere. Unfortunately, eq.~(\ref{eq:S_def}) is not particularly useful for measuring the entropy in practice. Ways to determine $S$ will be discussed in the methods section below.

Finally, we also measure the minimal monomer-monomer distance
\begin{equation}
r_{\rm min}\coloneqq \min_{i\ne j}(|\mathbf{x}_i-\mathbf{x}_j|)
\label{eq:rmin_def}
\end{equation}
or its deviation from the monomer diameter $\rho=r_{\rm min}-d$. Further below we will show how $\rho$ is closely related to the entropy and can be used to measure it.

\subsection{Boundary conditions}

Although ring-like polymers are also frequently investigated the majority of research focuses on standard chains with two ends. Monomers close to one of the two termini typically experience a different environment than monomers in the center; the system can be considered to possess free boundary conditions (FBC). The resulting inhomogeneity is, however, sometimes undesirable. For instance, if the average of properties like bond-bond correlations or internal distances is considered, it is preferable that these quantities do not depend on the position in the chain the measurement is taken at. It would also be interesting to see how the magnitude of the corrections to scaling differ between homogeneous and inhomogeneous systems.

\begin{figure}
\begin{center}
\includegraphics[width=0.95\columnwidth]{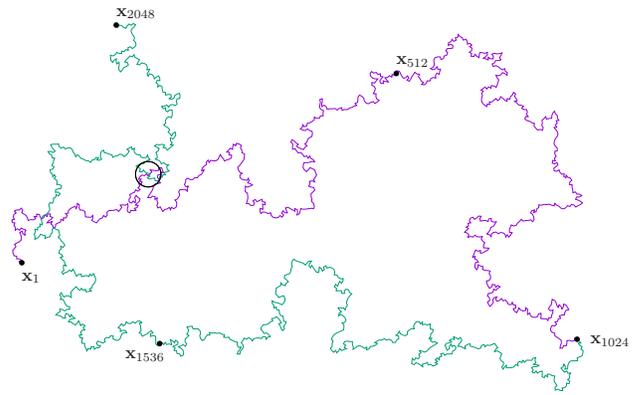}
\caption{\small{\label{fig:conf_pbc} A configuration and its  first image for $D=2$ and $L=1023$ with PBC. Although some monomers (e.g., $\mathbf x_{165}$ and $\mathbf x_{1867}$ in the marked circle) overlap, the configuration is valid since their separation along the chain exceeds the chain length ($1867-165>L$).}}
\end{center}
\end{figure}

We create a system with periodic boundary conditions (PBC) by taking a copy of the chain, rotating it and displacing it so that the image of $\mathbf x_1$ can be attached to $\mathbf x_N$ using a new bond $\mathbf b_N$ (Fig.~\ref{fig:conf_pbc}). For any such operation it is always possible to define a transformation $\mathcal T$
\begin{equation}
\mathbf{x}'=\mathcal T\mathbf x = \mathcal{R}_T\mathbf{x}+\mathbf{d}_T,
\label{eqn:trans_pbc}
\end{equation}
where $\mathcal{R}_T$ is a rotation matix and $\mathbf{d}_T$ a displacement vector and which fulfills 
\begin{equation}
|\mathcal T\mathbf x_1 - \mathbf x_N| = b.
\end{equation}
We can write
\begin{equation}
\mathbf x'_{i}\equiv\mathbf x_{N+i} = \mathcal T\mathbf x_i.
\end{equation}
The condition in eq.~(\ref{eqn:excl_vol}) needs to modified for use with PBC. We choose
\begin{equation}
|\mathbf{x}_i-\mathbf{x}_j| \ge d \quad \text{if} \quad 0<|i-k|<N,
\label{eqn:excl_vol_pbc}
\end{equation}
meaning that a monomer only ``sees'' $N-1$ other monomers in either direction. It may overlap with its own copy and with monomers that are further away along the chain. This condition is equivalent to demanding that any segment of $N$ adjacent monomers in the chain does not overlap with itself.

To motivate the designation ``periodic boundary conditions'' it is convenient to imagine the polymer configuration expressed not in Cartesian coordinates, but rather by bond and torsion angles $(\beta_i,\tau_i)$ as pairs of local coordinates at each joint specifying the position of the next monomer based on the three previous ones: 
\begin{equation}
\mathbf x_{i+1}=\chi(\beta_i,\tau_i,\mathbf x_{i},\mathbf x_{i-1},\mathbf x_{i-2}).
\end{equation}
It is then not difficult to imagine that the chain is extended to infinity by simply repeating the existing sequence of pairs of angles over and over \footnote{Since a free-boundary polymer configuration with $N$ monomers only possesses $N-2$ bond angles and $N-3$ torsion angles, five additional angles need to be introduced.} leading to
\begin{eqnarray}
\mathbf x'_{1}\equiv\mathbf x_{N+1} &\coloneqq& \chi(\beta_N,\tau_N,\mathbf x_{N},\mathbf x_{N-1},\mathbf x_{N-2}),\\
\mathbf x'_{2}\equiv\mathbf x_{N+2} &\coloneqq& \chi(\beta_1,\tau_1,\mathbf x_{1},\mathbf x_{N},\mathbf x_{N-1}),\\
\mathbf x'_{3}\equiv\mathbf x_{N+3} &\coloneqq& \chi(\beta_2,\tau_2,\mathbf x_{2},\mathbf x_{1},\mathbf x_{N}),
\end{eqnarray}
etc.

In the past similar ideas have been used for theoretical considerations and periodic walks have been enumerated on lattices \cite{EndlessSAW}, however, to our knowledge in this study such boundary conditions are applied for the first time in the context of MC simulations.

\section{Methods}
\label{sec-Meth}

\subsection{Conformational updates and transformations}
In order to explore the state space of the polymers we perform MC simulations. To modify the configuration we use the pivot move \cite{Sokal}, which randomly selects a monomer $\mathbf x_k$ as a fulcrum and rotates all monomers on one side around it, e.g.:
\begin{equation}
\mathbf{x}'_i=\mathcal T_p\mathbf x_i = \mathcal{R}_p\mathbf{x}_i+\mathbf{d}_p \qquad \text{for} \quad i>k,
\label{eqn:pivot}
\end{equation}
with a random rotation matrix $\mathcal{R}_p$ and the vector $\mathbf{d}_p=\mathbf x_k-\mathcal{R}_p\mathbf x_k$. Of course, the update is only accepted if none of the moved monomers overlaps with the ones that keep their position. The side to be rotated can be freely chosen, due to the overall rotational symmetry of the system. We always pick the smaller, i.e., $\mathbf x_1,\dots,\mathbf x_{k-1}$ if $k\le N/2$ and $\mathbf x_{k+1},\dots,\mathbf x_N$ otherwise.

We also implemented a bond-rotation update, although its application was only necessary for $D=2$. We randomly select a bond vector $\mathbf b_k=|\mathbf x_{k+1}-\mathbf x_k|$ and assign a new direction to it, $\mathbf b_k\rightarrow \mathbf b'_k$. Since this implies
\begin{equation}
\mathbf{x}'_i = \mathbf x_i -\mathbf b_k +\mathbf b'_k  \qquad \text{for} \quad i>k,
\label{eqn:bondrot}
\end{equation}
the update can also trivially be expressed in the form of a transformation of the type used in eqs.~(\ref{eqn:trans_pbc}) and (\ref{eqn:pivot}):
\begin{equation}
\mathbf{x}'_i=\mathcal T_b\mathbf x_i = \mathcal{I}\mathbf{x}_i+\mathbf{d}_b \qquad \text{for} \quad i>k,
\end{equation}
with the identity matrix $\mathcal{I}$ and $\mathbf{d}_b=\mathbf b'_k-\mathbf b_k$.

\subsection{Binary tree method}

\begin{figure}
\begin{center}
\begin{minipage}{0.5\textwidth}
\begin{scaletikzpicturetowidth}{0.9\textwidth}
\begin{tikzpicture}[scale=\tikzscale,level 1/.style={sibling distance=16em},
		level 2/.style={sibling distance=8em},
		level 3/.style={sibling distance=4em},
		level 4/.style={sibling distance=2em},
		every node/.style={transform shape}]

\node(R) [circle,draw,fill=red!20] {1\dots16}
    child { node [circle,draw,fill=blue!20] {1\dots8} 
	  child { node [circle,draw,fill=green!20] {1\dots4} 
	    child { node [circle,draw,fill=gray!20] {1,2} 
	        child { node [circle,draw,fill=yellow!20] {1} 
	        }
	        child { node [circle,draw,fill=yellow!20] {2} }
	    }
	    child { node [circle,draw,fill=gray!20] {3,4} 
	        child { node [circle,draw,fill=yellow!20] {3} 
	        }
	        child { node [circle,draw,fill=yellow!20] {4} }
	    }
	  }
	  child { node [circle,draw,fill=green!20] {5\dots8} 
	    child { node [circle,draw,fill=gray!20] {5,6} 
	        child { node [circle,draw,fill=yellow!20] {5} 
	        }
	        child { node [circle,draw,fill=yellow!20] {6} }
	    }
	    child { node [circle,draw,fill=gray!20] {7,8} 
	        child { node [circle,draw,fill=yellow!20] {7} 
	        }
	        child { node [circle,draw,fill=yellow!20] {8} }
	    }
	  }
    }
    child { node [circle,draw,fill=blue!20] {9\dots16} 
	  child { node [circle,draw,fill=green!20] {9\dots12} 
	    child { node [circle,draw,fill=gray!20] {9,10} 
	        child { node [circle,draw,fill=yellow!20] {9} 
	        }
	        child { node [circle,draw,fill=yellow!20] {10} }
	    }
	    child { node [circle,draw,fill=gray!20] {11,12} 
	        child { node [circle,draw,fill=yellow!20] {11} 
	        }
	        child { node [circle,draw,fill=yellow!20] {12} }
	    }
	  }
	  child { node [circle,draw,fill=green!20] {13\dots16} 
	    child { node [circle,draw,fill=gray!20] {13,14} 
	        child { node [circle,draw,fill=yellow!20] {13} 
	        }
	        child { node [circle,draw,fill=yellow!20] {14} }
	    }
	    child { node [circle,draw,fill=gray!20] {15,16} 
	        child { node [circle,draw,fill=yellow!20] {15} 
	        }
	        child { node [circle,draw,fill=yellow!20] {16} }
	    }
	  }
    }
;

\node [left=4cm of R] {\huge{(a)}};

\end{tikzpicture}
\end{scaletikzpicturetowidth}
\vspace{.5cm}
\includegraphics[width=0.9\textwidth]{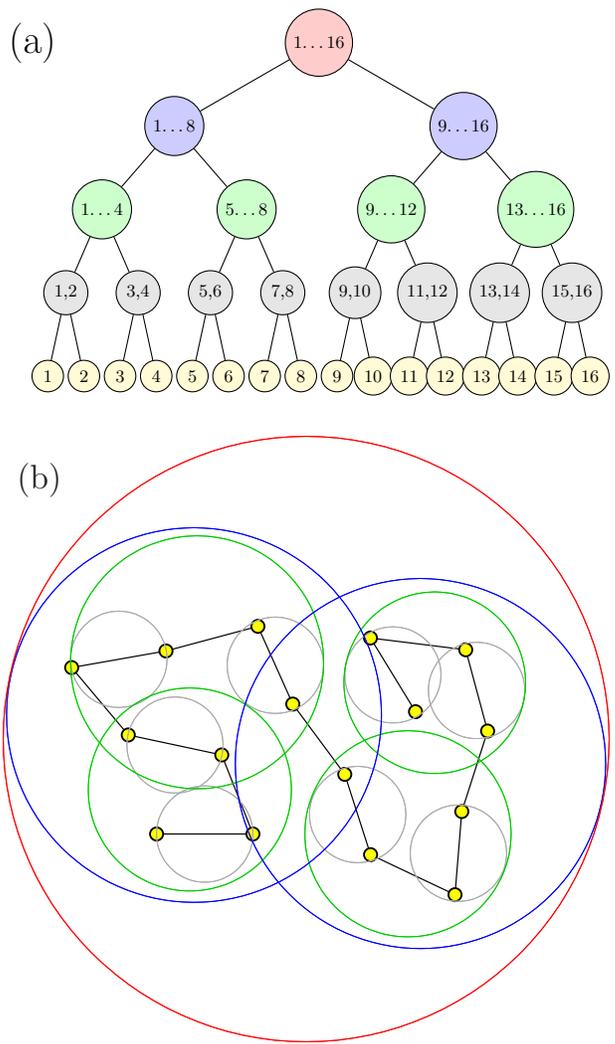}
\caption{\small{\label{fig:conf_tree} In the binary tree (a) each node represents a group of monomers and stores the parameters of a sphere (b) containing them. (Although individual monomers are drawn as yellow circles with finite size, the respective spheres have radii of zero.)}}
\end{minipage}
\end{center}
\end{figure}

A few years ago Clisby \cite{Clisby1,Clisby2,Clisby3} introduced a powerful new method for the simulation of self-avoiding walks. In our recent publication \cite{our_cpc2020} we adapted this technique in order to be able to investigate the system at hand. Without it, chains with $L\approx10^6$ as considered in this study could not be simulated. Here, we will only give a brief overview of the main elements of the algorithm. For details we refer to \cite{our_cpc2020}. Central to the method is a binary tree whose leaves correspond to individual monomers and inner nodes store collective information describing all monomers of the subtree to which they are root. This information contains two essential parts:

First, the node stores the parameters of a sphere that contains all monomers in the subtree (Fig.~\ref{fig:conf_tree}). This allows one to test whether the distance of two nodes' spheres exceeds $d$. If it does none of the monomers in one node can possibly overlap with a monomer in the other, while a distance smaller than $d$ or intersection of the spheres leaves the matter undecided. In the latter case one proceeds by replacing one node by its children and tests for the resulting two pairs of spheres. In this manner a recursive algorithm that determines whether monomers from two distinct groups overlap can be implemented.

The second important element that is stored in a node is a transformation representing the aggregate of all MC moves that are yet to be applied to all nodes in the subtree. As one main design feature of the algorithm the polymer data is always accessed at a level as coarse grained, i.e., as high up the tree, as possible. This strategy was followed above reading the polymer's configuration while testing for overlaps, but it can also be employed while actively modifying the state of the chain. When a group of monomers is to be moved, instead of updating every monomer's position, only the sphere in the highest possible node is moved and the respective transformation stored in that node. It is applied to lower levels in the tree only when this becomes necessary, i.e., if tests for overlap can not be resolved higher up and coordinates from these nodes need to be accessed. If the node in which a transformation $\mathcal T_a$ is to be stored already contains a transformation $\mathcal T_b$, it is easily possible to calcuate their product $\mathcal T_c= \mathcal T_a \circ \mathcal T_b$ and store it instead.

It is convenient to store and update additional data in the nodes that are not required by the algorithm but facilitate measurements during the simulation. The center of mass $\mathbf c_a=\sum_{i=k}^{k+n_a-1}\mathbf x_i/n_a$ of a node's $n_a$ monomers $\mathbf x_k,\dots,\mathbf x_{k+n_a-1}$ is a natural choice. If it is available for the children of a node it can easily be calculated for the node itself. We also track for each node the squared radius of gyration $g_a^2=\sum_{i=k}^{k+n_a-1}(\mathbf x_i - \mathbf c_a)^2/n_a$. Again the goal is to recursively calculate $g$ for increasingly large groups of monomers  \cite{Clisby3}: If $g_l,g_r$ are known for the children $l$ and $r$ of a parent node $p$ it is
\begin{equation}
g_p^2=( n_l( g_l^2 + (\mathbf c_l - \mathbf c_p)^2) + n_r(g_r^2 + (\mathbf c_r - \mathbf c_p)^2) ) / n_p.
\end{equation}

In the following we will discuss simulations where the diameter $d$ is allowed to vary at run-time in order to measure the entropy. To judge whether an increase by a certain amount is possible it is required to keep track of the minimal monomer-monomer distance $r_{\rm min}$ [eq.~(\ref{eq:rmin_def})]  for the entire chain. This can also be facilitated by storing the respective value for each node. Now, however, obtaining the value for a parent node is more complicated, albeit thanks to the binary tree still reasonably efficient. In addition to the values from the child nodes it is required to determine the minimum distance between the two groups.

\subsection{Measuring entropy}

For self-avoiding walks there are two strategies for measuring the number of possible walks $c_L=e^S$. For smaller $L$ all walks can be enumerated. For $D=3$ the state-of-the-art is $L=36$ \cite{saw_enum_D3} while for $D=2$ in Ref.~\cite{saw_enum_D2_B} the author improved their own record \cite{saw_enum_D2} from $L=71$ to $L=79$. For longer walks one simulates two smaller walks of length $L$ independently and every now and then tests whether the two walks connected by a random bond vector produce a valid self-avoiding walk of length $2L+1$ \cite{Clisby_gamma,JointCut}. If the combination is valid with probability $p_{\rm c}(L)$ than 
\begin{equation}
c_{2L+1}=p_{\rm c}c_{L}^2z,
\label{eq:c2L}
\end{equation}
where $z$ is the coordination number of the lattice, i.e., the number of choices for the connecting vector. 

The second method can be used for our model as well and the results suggested a non-trivial dependence of the entropy $S$ on $d$, i.e., on the strength of the excluded-volume interaction. To study this dependency in more detail we introduce a new simulation technique, that allows the system to change the monomer diameter $d$ during the simulation, while fulfilling detailed balance. The goal is to determine the relative statistical weight of the set of all configurations for different $d$ and in particular relating it to the known value $S=L\ln \cal{S}_D$ for $d=0$. Just like with standard flat-histogram MC methods \cite{muca2,wl} we introduce a weight function $W(d)$ that defines the acceptance probability for a suggested change $d\rightarrow d'$. Provided that $d'-d<\rho$, i.e., that the new configuration is valid, the update is accepted with probability $P_{\rm acc}(d\rightarrow d')=\min(1,W(d')/W(d))$. The goal is a constant (flat) distribution of samples as a function of $d$ meaning that $\ln W(d)=-S(d)$. If the argument of the weight function is continuous as it is here, the interval is typically divided into sub-intervals of equal width on each of which $W$ is constant (binning). We decided to follow a different approach and allow only distinct values of the diameter $d\in\{0,h,2h,\dots,1-h,1\}$ where we chose $h=1/1024$. For chains longer than those we simulate with this method in this study it will become necessary to use smaller values. This discretization is an option because we have direct control over the possible changes of $d$, while in standard MC simulation changes of the respective quantity -- the energy -- can only be controlled indirectly.

The standard method of obtaining a flat distribution involves either an iterative process of several simulations where the produced histograms are used to successively improve $W(d)$ or a Wang-Landau-like procedure that constantly modifies $W$ with the changes becoming smaller over time. In our case, however, we can measure the probability that an increase in diameter is at all possible, i.e., the probability $P_{kh}(\rho>h)$ that $\rho>h$. Since given that $S((k+1)h)<S(kh)$ implies $W((k+1)h)>W(kh)$ such a change should always be accepted, detailed balance demands that for a flat distribution the inverse move should be accepted with the same probability. Therefore
\begin{equation}
\frac{W((k+1)h)}{W(kh)} = P_{\rm acc}((k+1)h\rightarrow kh) = P_{kh}(\rho>h).
\end{equation}

It turns out that measuring $$P_{kh}(\rho>h)$$ and updating $W$ simultaneously does allow the simulation to sample all values of $d$. It should be noted that by adjusting $W$ at runtime detailed balance is violated and in order to avoid systematic errors it is good practice to perform a final production run with fixed $W$ throughout which all data used in the analysis are obtained.

\subsection{Derivatives of entropy}

In the previous subsection it has become clear that the quantity $\rho$, the maximal amount by which the diameter $d$ can be increased for a given configuration, is intimately related to the entropy $S$ of the system. This relationship can be exploited to gain additional useful information. As we will show now, there are simple equations that link the moments of the distribution of $\rho$ to the derivatives of $S(d)$.

It is trivial that for any configuration the monomer diameter $d$ can always be reduced without creating any violation of eq.~(\ref{eqn:excl_vol}), i.e., no overlaps of monomers. However, increasing $d$ is not always allowed and the probability of a possible increase from $d_0$ to $d_0+\epsilon$ is
\begin{equation}
P_{d_0\rightarrow d_0+\epsilon}=P_{d_0}(\rho>\epsilon)=e^{S(d_0+\epsilon)-S(d_0)}.
\end{equation}

In other words $e^{S(d_0+\epsilon)-S(d_0)}$ denotes the fraction of configurations for $d_0$ that are still valid for $d_0+\epsilon$ or for which $\rho>\epsilon$. For very long chains the possible changes will become very small $\epsilon\ll 1$ and the distribution for $\rho$ at $d_0$ which we denote $p_{d_0}(\rho)$ will be very similar to the distribution for $\rho$ at $d_0+\epsilon$ since $d_0$ and $d_0+\epsilon$ are very similar. It is, therefore, justified to use the Ansatz
\begin{equation}
p_{d_0}(\rho)\propto e^{\beta \rho + \gamma \rho^2 + \dots}.
\end{equation}
Taking the derivative with respect to $\epsilon$ of
\begin{equation}
P_{d_0}(\rho\ge\epsilon)=\int_\epsilon^\infty p_{d_0}(\rho) \diff \rho=e^{S(d_0+\epsilon)-S(d_0)}
\end{equation}
we get
\begin{equation}
-p_{d_0}(\epsilon) = S'(d_0+\epsilon)e^{S(d_0+\epsilon)-S(d_0)}
\end{equation}
and
\begin{widetext}
\begin{eqnarray}
-e^{\beta \epsilon + \gamma \epsilon^2 + \dots} &\propto& S'(d_0+\epsilon)e^{S(d_0)+S'(d_0)\epsilon+\frac12S''(d_0)\epsilon^2+\dots-S(d_0)},\\ \nonumber
                                  &=& S'(d_0+\epsilon)e^{S'(d_0)\epsilon+\frac12S''(d_0)\epsilon^2+\dots},\\
                                  &=& S'(d_0)\left(1+\frac{S''(d_0)}{S'(d_0)}\epsilon +\frac{S'''(d_0)}{2S'(d_0)}\epsilon^2+\dots\right)e^{S'(d_0)\epsilon+\frac12S''(d_0)\epsilon^2+\dots}.
\end{eqnarray}
\end{widetext}
Since $S(\rho)$ and its derivatives diverge linearly with $L$ the ratios $S''/S',S'''/S',\dots$ converge and $S',S'',\dots$ can be identified with $\beta,2\gamma,\dots$. The latter can also be expressed in terms of moments of $\rho$ leading for long chains or small $\langle\rho\rangle$ to
\begin{equation}
\frac{\partial S}{\partial d}\approx-\frac{1}{\langle\rho\rangle}
\label{eq:S_deriv_I}
\end{equation}
and
\begin{equation}
\frac{\partial^2S}{{\partial d}^2}\approx \frac{\langle\rho^2\rangle-2\langle\rho\rangle^2}{2\langle\rho\rangle^4}
\label{eq:S_deriv_II}
\end{equation}
with equality in the limit $L\rightarrow\infty$.

\section{Results}
\label{sec-Res}

With the described method it is possible to simulate very long chains, although the efficiency depends strongly on the number of dimensions $D$ and the monomer diameter $d$. While for $D=2,3$ and not too small diameters chain length of $L\approx10^6$ are easily accessible, for $D=4,5$ or small diameters, i.e., if the behavior of the chain is close to a random walk, $L\approx10^4$ can already be challenging. We also simulated for $d=0$, i.e., the pure random walk, in order to verify that the algorithm produces correct results and that statistical errors are reasonable.

Since the behaviors of the end-to-end distance and the radius of gyration are very similar we are not going to treat the latter in detail here. Rather we focus mainly on the end-to-end distance and in Sec. \ref{sec-Res} D discuss the weakly universal ratio of the two length measures.

\subsection{Scaling and crossover of the end-to-end distance for $D=2,3$}

For $D<4$ the mean squared end-to-end distance of a self-avoiding walk on a lattice is expected to behave like
\begin{equation}
\langle R^2 \rangle = AL^{2\nu}\left(1+\frac{a_1}{L}+\frac{a_2}{L^2}+\dots+\frac{b_1}{L^{\Delta}}+\frac{b_2}{L^{2\Delta}}+\dots \right),
\label{eq:saw_scal}
\end{equation}
with additional correction terms arising from the lattice geometry. Here, $\nu$ is the Flory exponent and $\Delta$ the exponent of confluent corrections to scaling, both of which are universal. This scaling also applies to the hard-sphere polymer ($d>0$) in continuous space where lattice corrections are absent.

\begin{figure}
\begin{center}
\includegraphics[width=0.95\columnwidth]{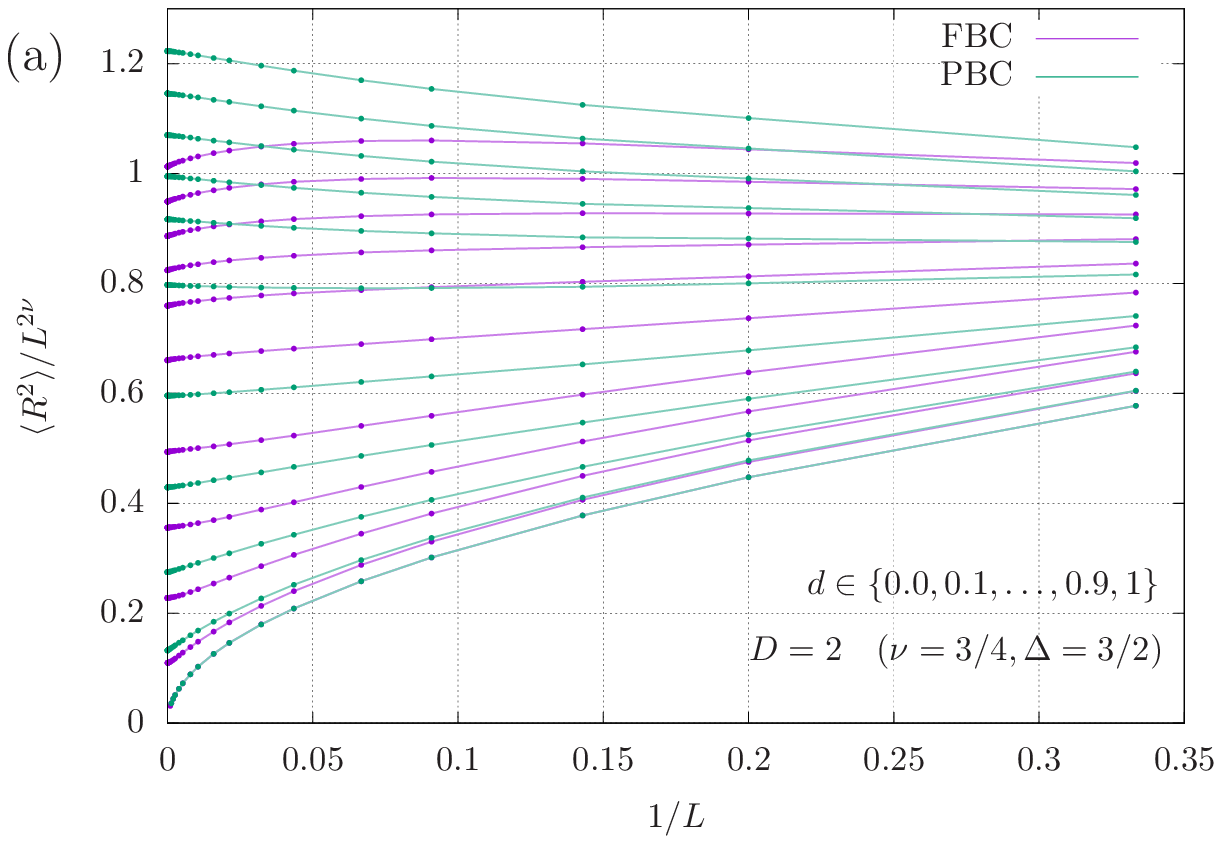}\\
\includegraphics[width=0.95\columnwidth]{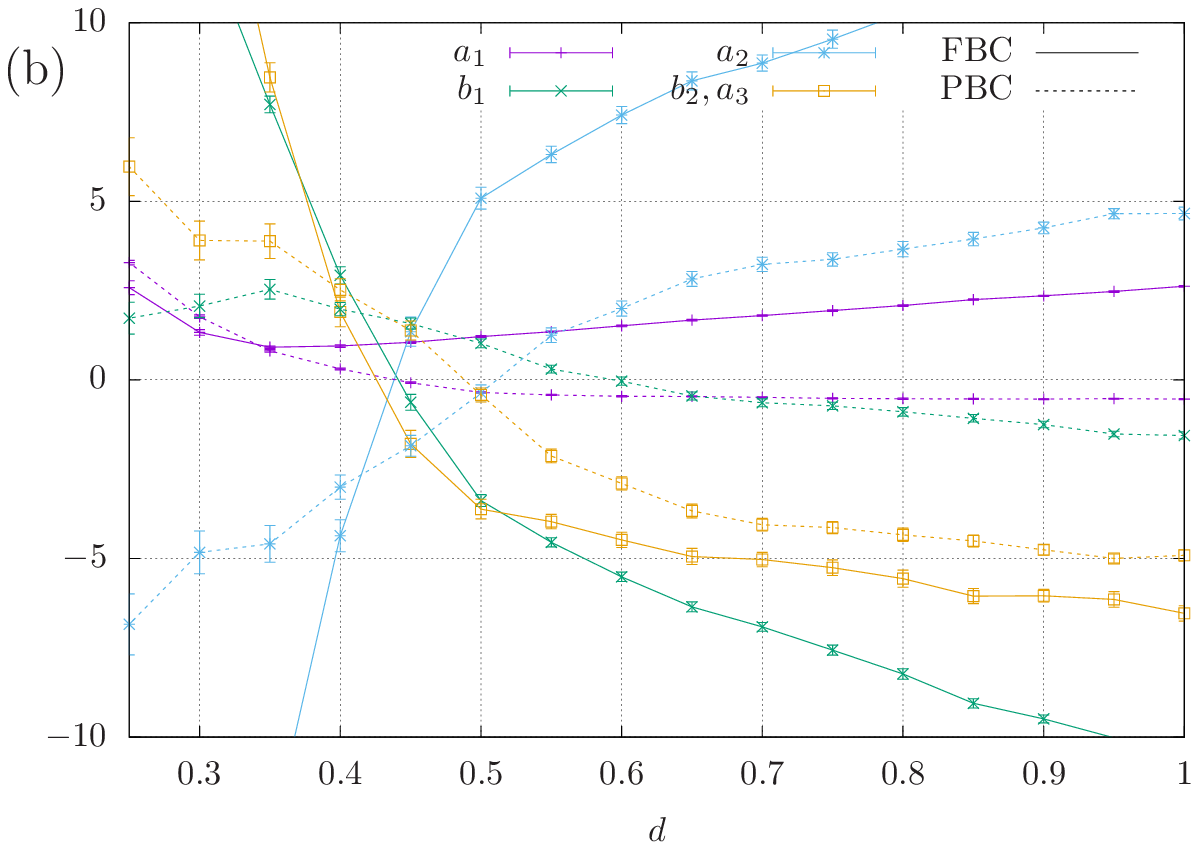}\\
\caption{\small{\label{fig:ree_dist_D2_B} (a) Scaling of end-to-end distance for $D=2$. The diameter $d$ decreases from top to bottom. Errors are smaller than the line width. (b) The coefficients of the correction terms for FBC (solid) and PBC (dashed) . Note that $L^{-3}=L^{-2\Delta}$ implies that the terms $a_3$ and $b_2$ cannot be distinguished.}}
\end{center}
\end{figure}

For $D=2$ the values of the exponents are known exactly: $\nu=3/4$ and $\Delta=\omega\nu=3/2$, which means that the leading correction is of order $L^{-1}$. In Fig.~\ref{fig:ree_dist_D2_B}(a) we show $\langle R^2 \rangle/L^{2\nu}$ for different diameters $d$ and $L\le 10^6$. Note that for the random walk ($d=0$) the choice of boundary conditions is irrelevant and the two curves coincide, while the linear scaling of $\langle R^2 \rangle$ with $L$ leads to $\langle R^2 \rangle/L^{2\nu}\propto L^{-1/2}$. In contrast, for $d>0$ the curves become linear for small $L^{-1}$ indicating that the leading correction is of this order. For larger diameters the boundaries have a dramatic influence not only on the amplitudes $A$; switching from FBC to PBC even changes the sign of the first correction. The coefficients of the first four correction terms obtained through fitting are shown in Fig.~\ref{fig:ree_dist_D2_B}(b). One interesting observation is that close to $d=0.45$ for FBC the higher-order corrections become very small, meaning that the curve in Fig.~\ref{fig:ree_dist_D2_B}(a) is close to a straight line. We also observe that for most values of $d$ the corrections to scaling for PBC are indeed smaller than for FBC.

\begin{figure}
\begin{center}
\includegraphics[width=0.95\columnwidth]{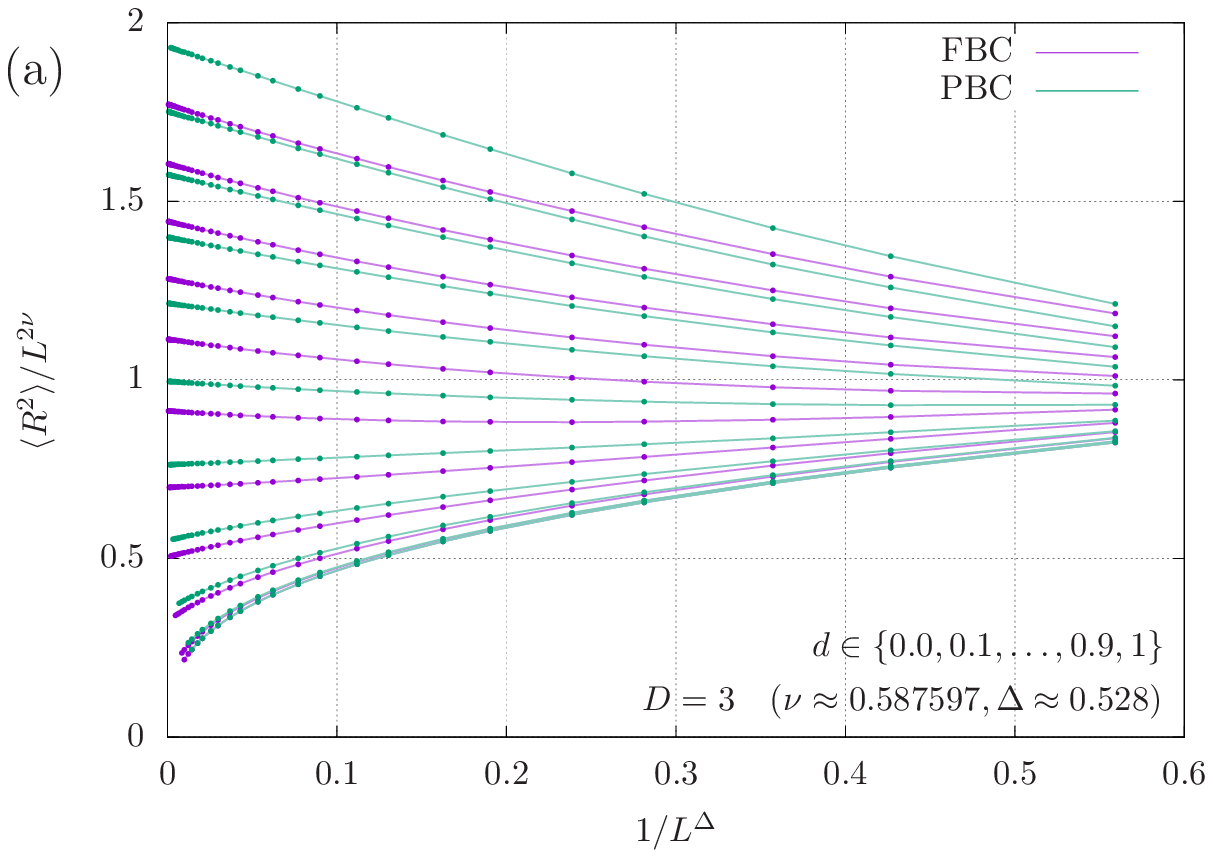}\\
\includegraphics[width=0.95\columnwidth]{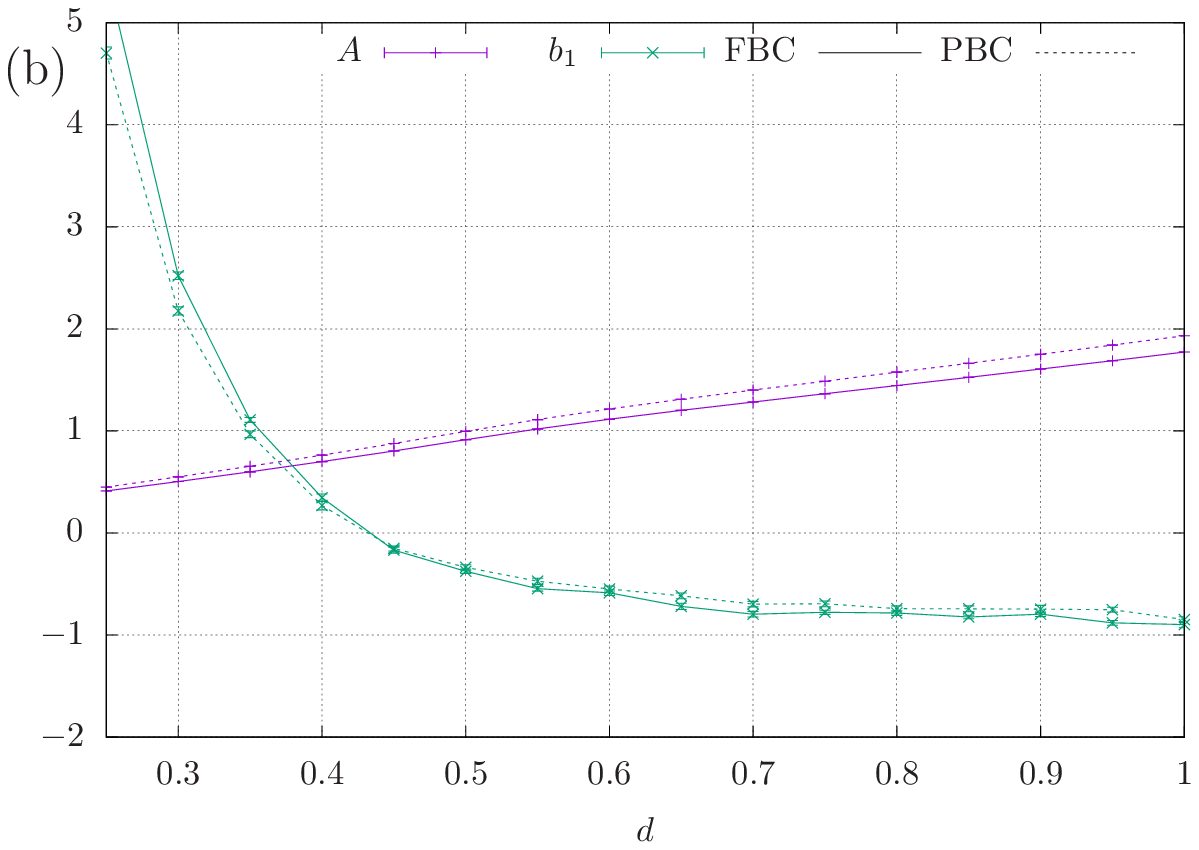}
\caption{\small{\label{fig:ree_dist_D3_B} (a) Scaling of end-to-end distance for $D=3$. The diameter $d$ decreases from top to bottom. (b) Amplitude $A$ and first correction $b_1$.}}
\end{center}
\end{figure}

For $D=3$ [Fig.~\ref{fig:ree_dist_D3_B}(a)] the dominant correction term for long chains is of order $L^{-\Delta}\approx L^{-0.5}$ and displaying $\langle R^2 \rangle/L^{2\nu}$ accordingly again leads to linear curves for large $L$ and  $d>0$. There is some uncertainty with regard to the value of $\Delta$. While field-theoretic methods predict values close to or slightly below $\omega=\Delta/\nu=0.85$ \cite{Kleinert,Shalaby}, implying $\Delta$ close to or slightly below $0.5$, recent MC simulations \cite{Clisby3} produced values of $\nu=0.58759700(40)$ and $\Delta=0.528(8)$ which means $\omega=0.90(1)$. For the analysis of our data we use the latter value, since it is closer to our own estimates. The values of the relevant universal exponents are compiled in Table \ref{tab:exponent}. Unfortunately, with $2\Delta\approx1$ there are two correction terms of similar but different order which makes it difficult to obtain the parameters $a_1$ and $b_2$. We only get good values for $b_1$ shown in Fig.~\ref{fig:ree_dist_D3_B}(b). We find again that $|b_1|$ is slightly smaller for PBC although the difference is marginal.

\begin{table}
\caption{\small{\label{tab:exponent} Values of the universal exponents $\nu, \gamma$, and $\Delta$.}}
\begin{tabular}{c|c|c|c}
  $D$ &  $\nu$     & $\gamma$ & $\Delta$ \\ \hline
  $2$ & $3/4$     & $43/32$ &      $3/2$ \\
  $3$ & $\approx0.5875970$ & $\approx1.156953$ &    $\approx0.528$ \\
  $4$ & $0.5$ &   $1$ & \\
  $5$ & $0.5$ &   $1$ & \\
\end{tabular}
\end{table}

\begin{figure}
\begin{center}
\includegraphics[width=0.95\columnwidth]{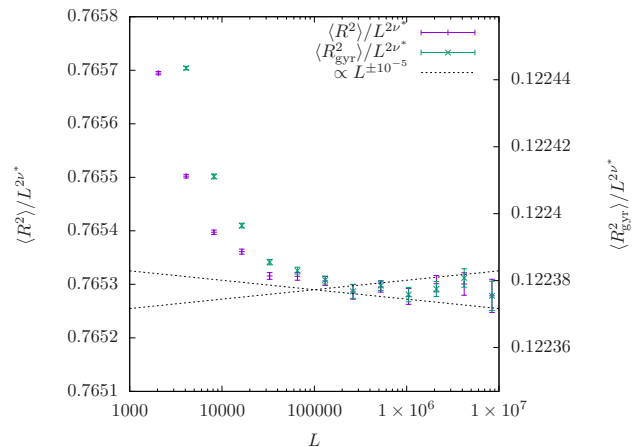}
\caption{\small{\label{fig:D3_nu} Squared end-to-end distance and radius of gyration divided by $L^{2\nu^*}$ in $D=3$ for $d=0.43225$ with $\nu^*=0.597604$.}}
\end{center}
\end{figure}

We also see in Fig.~\ref{fig:ree_dist_D3_B}(b) for both FBC and PBC and $D=3$ that $b_1$, i.e., the first correction to scaling vanishes for $d\approx0.43$. This can be exploited to determine the exponent $\nu$. We simulated particularly long chains up to $L\approx 10^7$ for $d=0.43225$. From fitting the data we obtain a value $\nu^*=0.587604(8)$ as shown in Fig.~\ref{fig:D3_nu}, where we also display the respective data for the radius of gyration.

\begin{figure}
\begin{center}
\includegraphics[width=0.95\columnwidth]{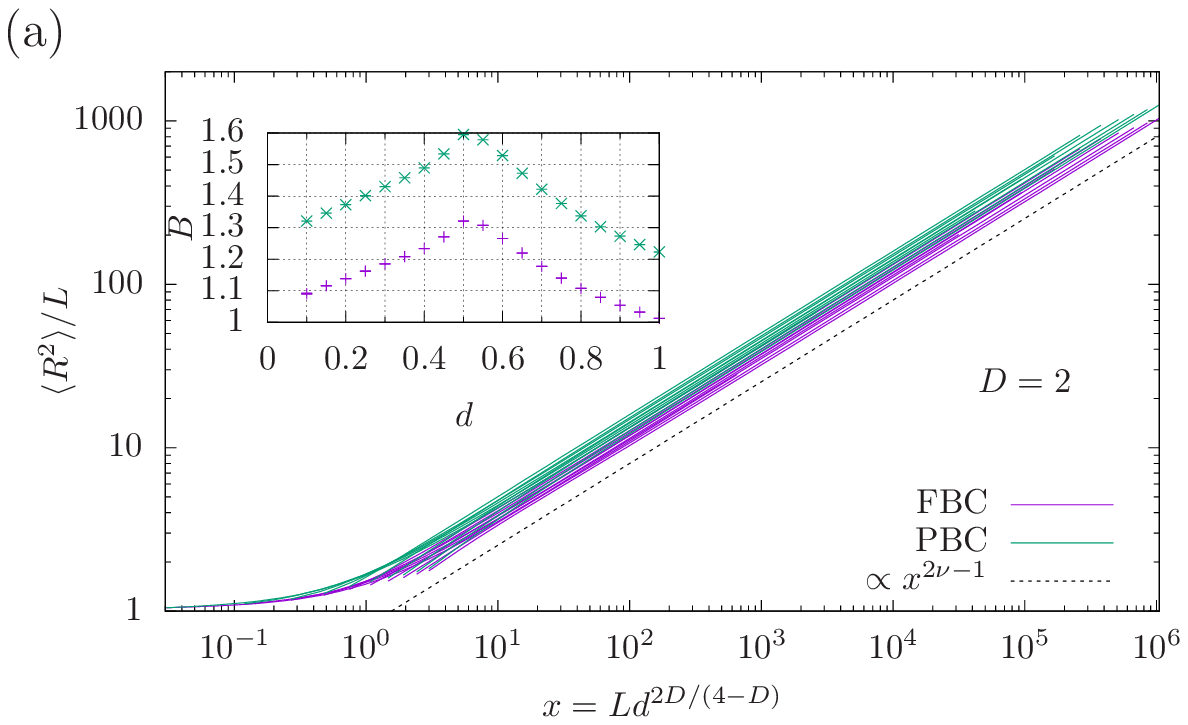}\\
\vspace{0.5cm}
\includegraphics[width=0.95\columnwidth]{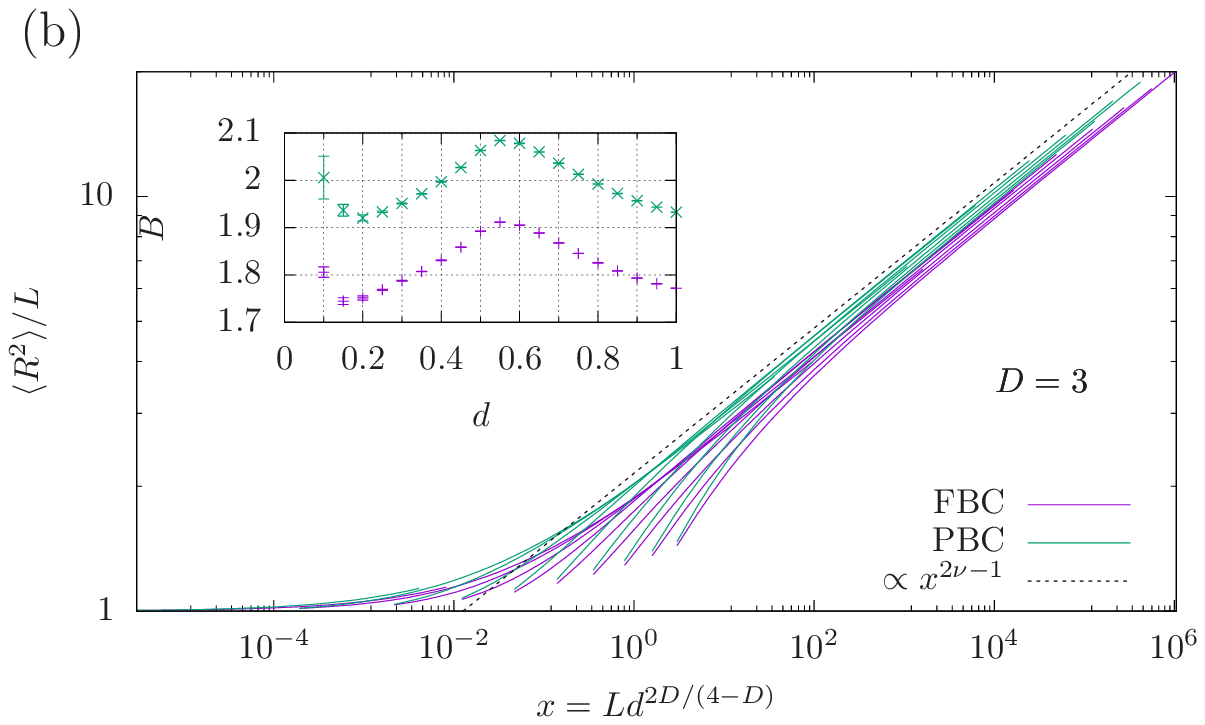}
\caption{\small{\label{fig:ree_co_D23} Crossover scaling for $D=2$ (a) and $D=3$ (b). The insets show the amplitudes $B$ of the asymptotic scaling. For $D=3$ and small diameters $d$ it becomes difficult to reliably determine $B(d)$ through fitting since $x\ll1$ for all available values of $L$.}}
\end{center}
\end{figure}

\begin{figure}
\begin{center}
\includegraphics[width=0.95\columnwidth]{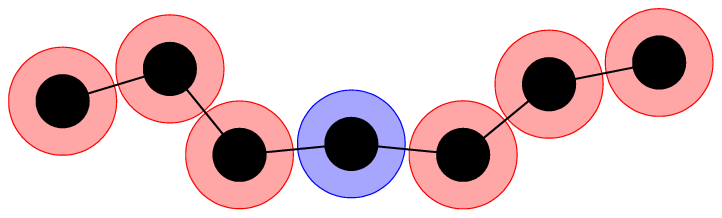}\\
\vspace{0.5cm}
\includegraphics[width=0.95\columnwidth]{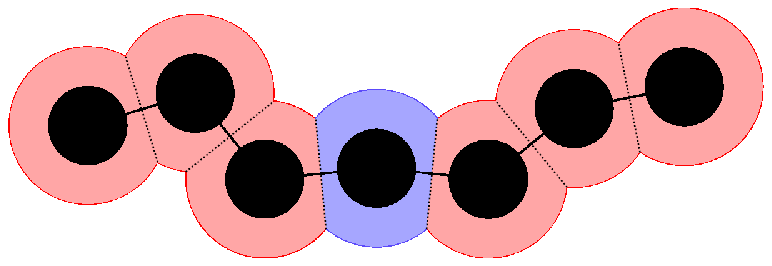}
\caption{\small{\label{fig:excl_vol} The excluded volume (red/blue) around the monomer chain (black) that is inaccessible for other parts of the chain. While if $d\le0.5$ (top) this volume for each monomer is a sphere and therefore proportional to $d^D$, for $d>0.5$ (bottom) the spheres overlap and the excluded volume per monomer grows more slowly with $d$.}}
\end{center}
\end{figure}

The above representation notwithstanding for long chains $\langle R^2 \rangle$ should not depend on two parameters -- the chain length and the range of repulsion -- but on a combination of the two. Using the bead diameter $d$ for the latter it should be
\begin{equation}
\langle R^2 \rangle/L=f(Ld^{D/\phi})
\label{eq:crosover_scal}
\end{equation}
in the limit of large $L$ with the crossover exponent $\phi=(4-D)/2$. Since eventually eq.~(\ref{eq:saw_scal}) should be valid for all $d>0$ it follows that $f(x)=Bx^{2\nu-1}$ for $x\rightarrow\infty$. As can be seen in Fig.~\ref{fig:ree_co_D23} this description is generally valid for $D=2$ and $3$ and the exponents describe the asymptotic scaling correctly. However, the data do not collapse perfectly and we observe a variation of $B$ depending on both the boundary conditions which is expected, but surprisingly also on the monomer diameter $d$ with a maximum in the proximity of $d=0.5$. Assuming that this is a direct result of the non-trivial way the excluded volume depends on $d$ (see Fig.~\ref{fig:excl_vol}) we used a parameter $\tilde{d}$ proportional to the square root (for $D=2$) of the excluded volume per monomer instead of $d$, however, this did not lead to a constant $B(\tilde{d})$.

\subsection{End-to-end distance for $D=4$}
For $D=4$ the self-avoiding walk has the same exponent as the random walk $\nu=1/2$, but there are logarithmic corrections \cite{Duplantier} due to $D=4$ being the upper critical dimension:
\begin{equation}
\langle R^2\rangle/L = A\left[\ln(L/\lambda)\right]^{1/4}\left[1-\frac{17\ln(4\ln(L/\lambda))+31}{64\ln(L/\lambda)}+\dots\right].
\label{eq:saw_corr_D4}
\end{equation}

\begin{figure}
\begin{center}
\includegraphics[width=0.95\columnwidth]{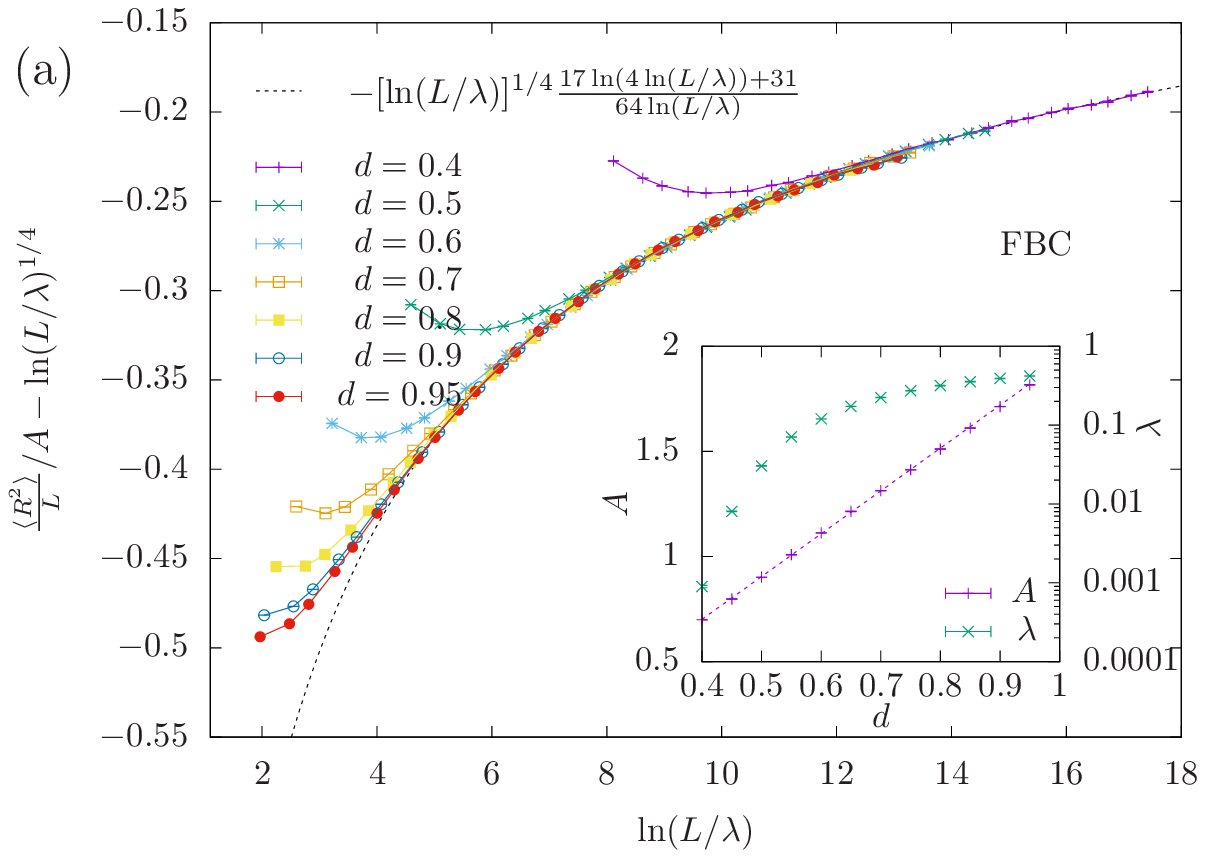}
\vspace{1.cm}
\includegraphics[width=0.95\columnwidth]{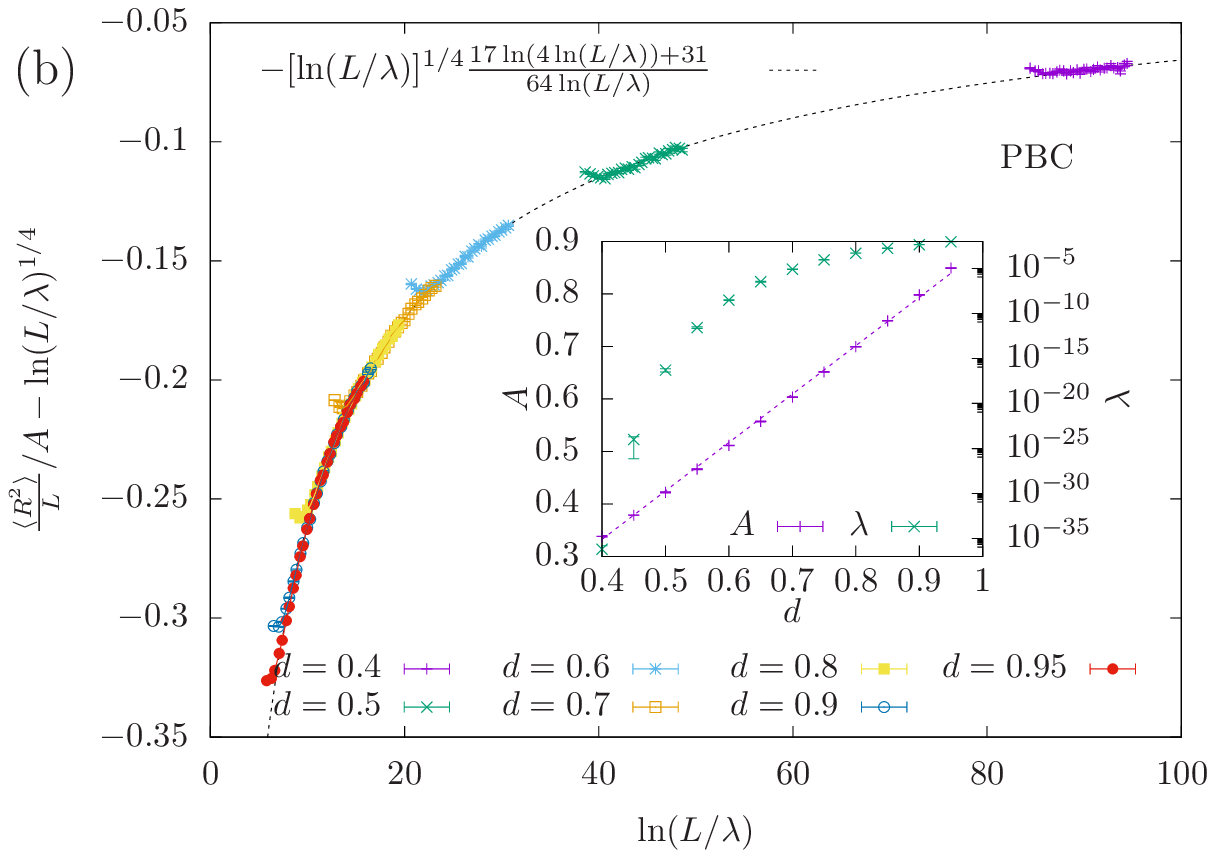}
\caption{\small{\label{fig:ree_dist_D4} Scaling of end-to-end distance for $D=4$ with FBC (a) and PBC (b).}}
\end{center}
\end{figure}

Previous studies have been performed on lattice walks with FBC \cite{saw_4D_Grassberger,saw_D4_Clisby} and although the exponent $1/4$ was not confirmed to complete satisfaction, agreement with the prediction is good if the walks are long enough ($>10^6$). Here, we managed to simulate chains with up to $L=3\times10^5$. Fitting the data for the longer chains led to values of $\lambda$ and $A$ displayed in the insets of Fig.~\ref{fig:ree_dist_D4} and to the rescaled data points in the main plots. Since the correction term in eq. (\ref{eq:saw_corr_D4}) has a fixed amplitude, it is $\lambda$ that specifies the magnitude of the correction corresponding to the curvature in the predicted curve in Fig.~\ref{fig:ree_dist_D4}. We observe that for larger $L$ the data collapse nicely, although higher-order corrections are manifest for smaller chains. There is a substantial difference between FBC and PBC. The values for $\lambda$ are larger for PBC by a factor of $\approx40$ for $d=0.95$ to many orders of magnitude for smaller diameters implying that the logarithmic corrections given in eq. (\ref{eq:saw_corr_D4}) are smaller for PBC. Furthermore, higher-order corrections indicated by deviations from the dashed line in Fig.~\ref{fig:ree_dist_D4} are smaller for PBC as well.
Field theory \cite{Kleinert} predicts $\nu^{-1}=2-\varepsilon/4+\dots$ for small $\varepsilon=4-D$. With eq. (\ref{eq:crosover_scal}) this suggests $A\propto d$ which is in good agreement with our results.

\subsection{End-to-end distance for $D=5$}

\begin{figure}
\begin{center}
\includegraphics[width=0.95\columnwidth]{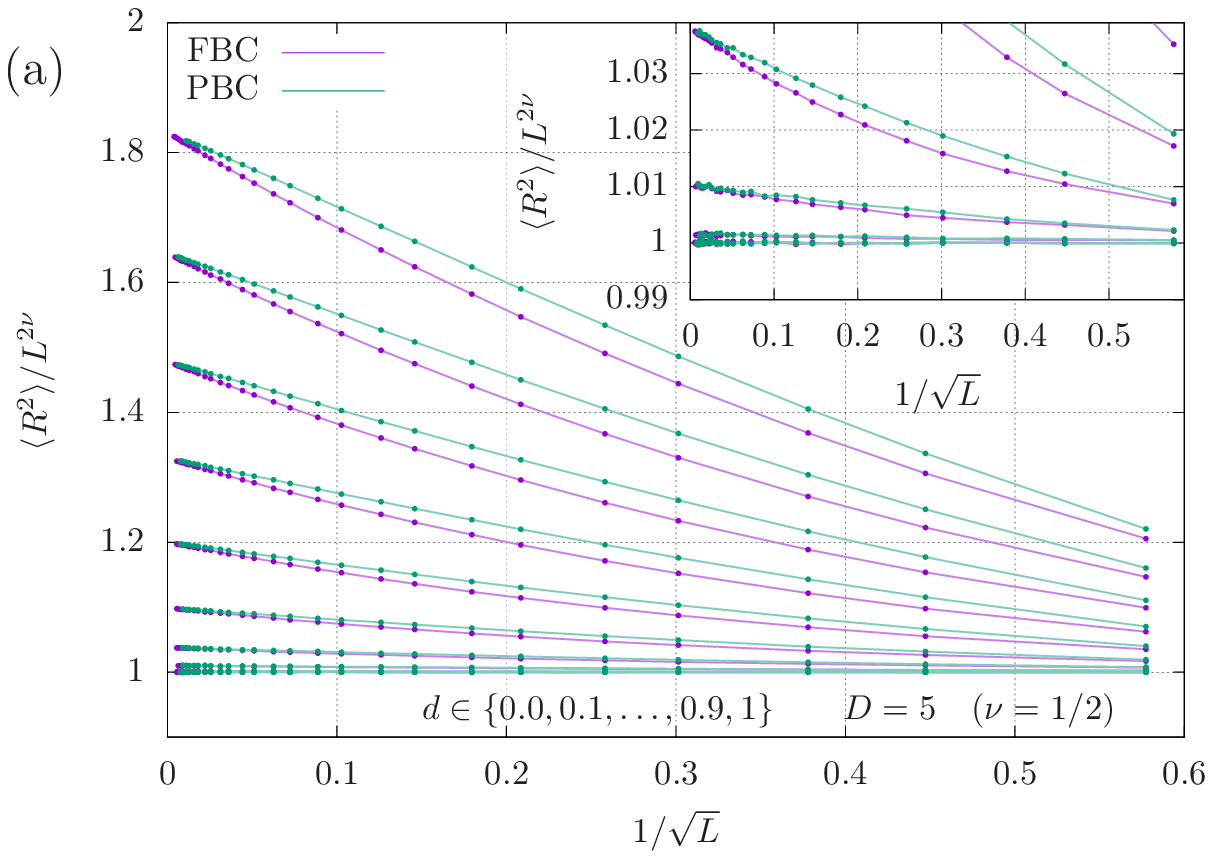}\\
\includegraphics[width=0.95\columnwidth]{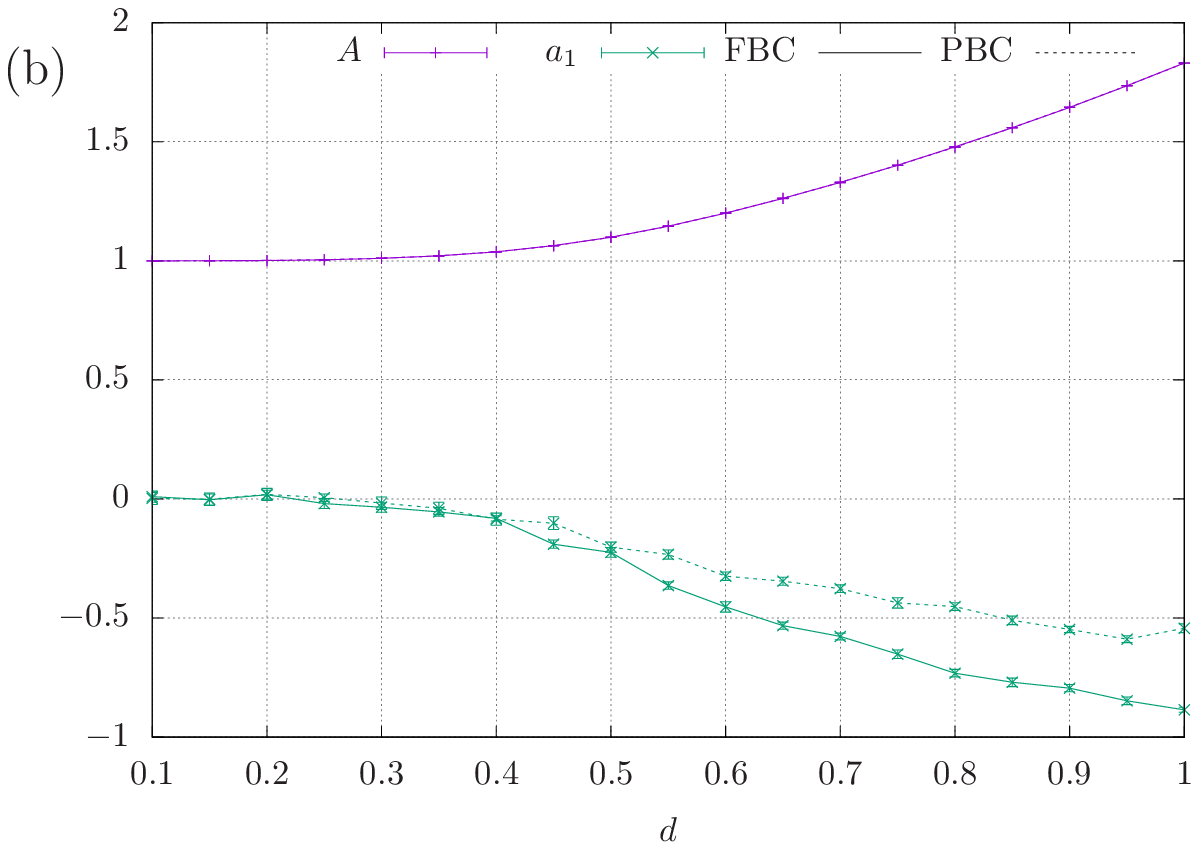}
\caption{\small{\label{fig:ree_dist_D5_B} (a) Scaling of end-to-end distance in five dimensions. Detail for small $d$ shown in the inset. The diameter $d$ decreases from top to bottom. Note that results for $d=0.1$ are within our statistical errors numerically and visually (see inset) indistinguishable from the pure random walk $d=0$. (b) Amplitude $A$ and coefficient $a_1$ of the first correction $\propto L^{-1/2}$.}}
\end{center}
\end{figure}

Above the upper critical dimension $D>4$ ``the puzzles of finite-size scaling are still not fully resolved'' \cite{RKenna}. Although there is only the one Gaussian fixed point it has been argued \cite{saw_D5_Prellberg} that the way excluded volume impacts scaling should still be quantifiable by the crossover exponent $\phi=(4-D)/2$ and that the leading-order correction to scaling for $D=5$ should, therefore, be of order $L^{-1/2}$. In the same study this conjecture is well supported by numerical data for the end-to-end distance of five-dimensional SAWs on a hypercubic lattice. Our data shown in Fig.~\ref{fig:ree_dist_D5_B} confirms this now also for hard-sphere polymers; we point out, however, that it is not possible to find a single scaling function as in eq.~(\ref{eq:crosover_scal}) since in the thermodynamic limit $\langle R^2 \rangle/L$ converges to different values for different diameters $d$.

\subsection{Amplitude ratios}
\begin{figure}
\begin{center}
\includegraphics[width=0.95\columnwidth]{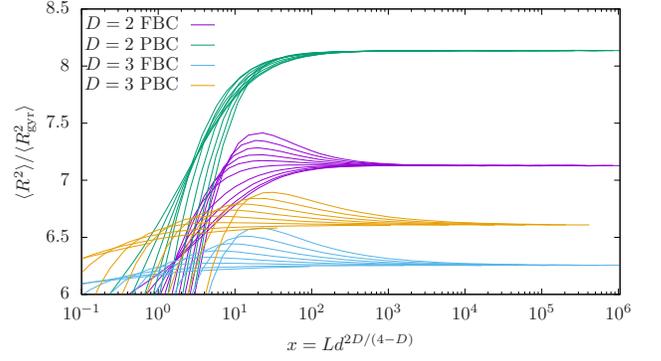}
\caption{\small{\label{fig:amp_rat} Amplitude ratios $\langle R^2\rangle / \langle R_{\rm gyr}^2\rangle$ for $D=2$ and $3$ and different diameters $d$.}}
\end{center}
\end{figure}

  We also want to discuss the ratio of amplitudes $g_D=A_{R}/A_{R_{\rm gyr}}$ corresponding to the thermodynamic limit $\lim_{L\rightarrow\infty}(\langle R^2\rangle_L / \langle R_{\rm gyr}^2\rangle_L)$. They are thought of as being of weak universality: they depend on dimensionality, but in the case of SAW do not change with the lattice type. They are, however, different for linear walks and closed polygons. From our simulations for $D=2$ and $D=3$ we obtain the amplitude ratios shown in Fig.~\ref{fig:amp_rat} whose asymptotic limits for large $L$ are listed in Table \ref{tab:ratio} and agree with values from the literature on SAWs, e.g., $g_2=7.129(4)$ \cite{Li_Madras_Sokal} and $g_3=6.25353(1)$ \cite{Clisby3}. Our value for $g_3$ with FBC was obtained from the simulations of particularly long chains up to $L\approx10^7$ with $d=0.43225$ where the correction of lowest order is close to zero. Therefore, the statistical uncertainty is smaller for this value. For PBC we observe larger values for the ratio, once more showing that $g_D$ is only weakly universal. Regardless of the number of dimensions for the random walk it is $g_D=6$ which consequently is also the limiting ratio we observe for the hard-sphere polymer with $D=4$ and $D=5$. In general the difference $g_D-6$ can be considered a quantitative measure for how strongly the chain or walk deviates from the random walk. It is no surprise that we observe a greater value for PBC, since for FBC both ends of the chain experience a less crowded environment than the center (or any part of the chain for PBC) and they are, therefore, less affected by the excluded-volume repulsion.

\begin{table}
\caption{\small{\label{tab:ratio} Ratio of amplitudes $g_D$ for different $D$ and boundary conditions.}}
\begin{tabular}{c|c|c}
  $D$ &  FBC     & PBC \\ \hline
  $2$ & $7.1278(2)$ & $8.1356(2)$  \\
  $3$ & $6.25352(2)$ & $6.606(1)$  \\
\end{tabular}
\end{table}

\subsection{Entropy}

\begin{figure}
\begin{center}
\includegraphics[width=0.95\columnwidth]{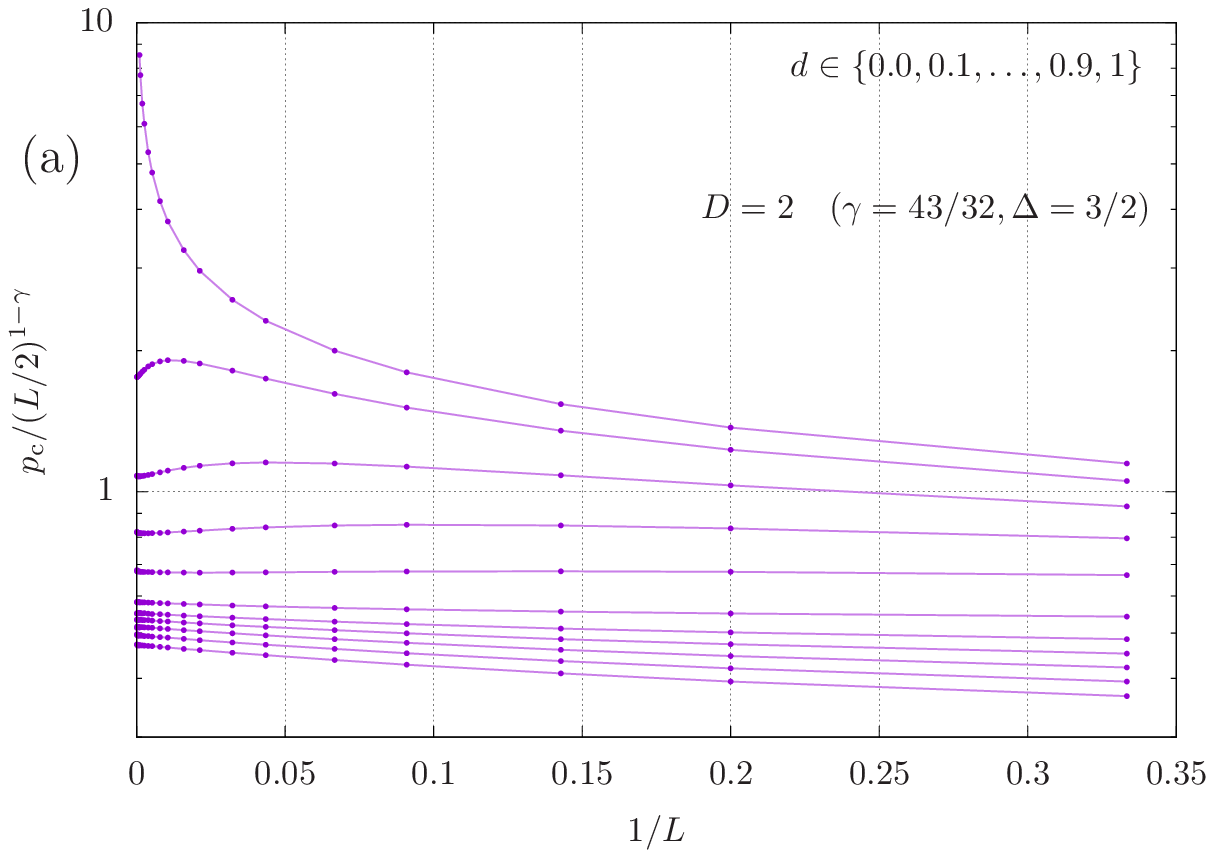}\\
\includegraphics[width=0.95\columnwidth]{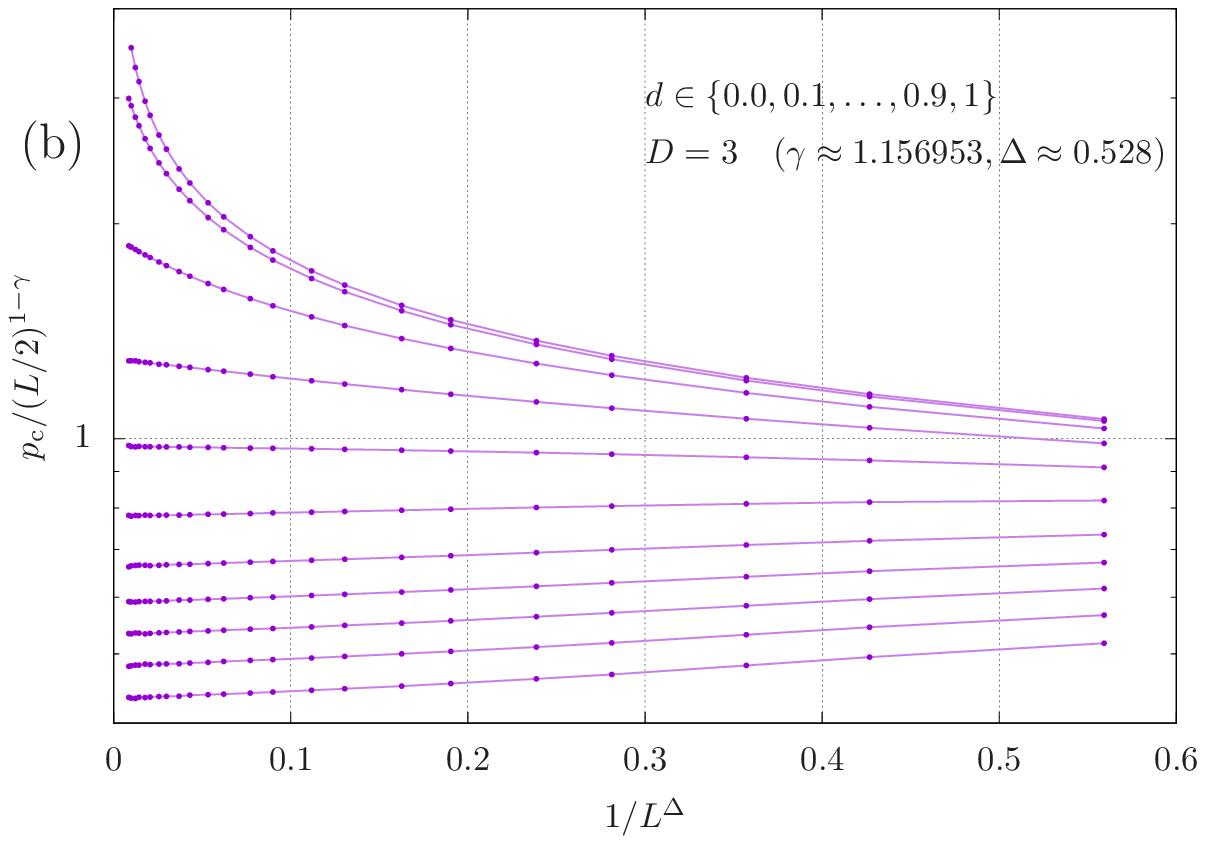}
\caption{\small{\label{fig:connect_D23} Probability $p_{\rm c}$ to successfully connect two chains of length $L$ divided by $(L/2)^{1-\gamma}$ as function of $L^{-\Delta}$ for $D=2$ (a) and $D=3$ (b) with FBC. Here the diameter $d$ \emph{increases} from top to bottom. Except for the pure random walk $(d=0)$ the curves converge for $L^{-\Delta}\rightarrow0$, confirming the expected scaling.}}
\end{center}
\end{figure}

As discussed earlier, one method to evaluate the entropy of a system with FBC is to measure the probability $p_{\rm c}(L)$ that two independent chains of length $L$ can be connected to form a non-overlapping chain of length $2L+1$. 
This probability relates to the volume of accessible state space $e^{S(L)}$ according to
\begin{equation}
p_{\rm c}(L)=\frac{e^{S(2L+1)}}{{\cal S}_D e^{2S(L)}},
\end{equation}
where we replaced in eq. (\ref{eq:c2L}) the coordination number $z$ by the surface of the $D$-sphere ${\cal S}_D$.
Assuming that the scaling law for the number of SAWs $c_L$ describes the behavior of $e^{S(L)}$ as well
\begin{equation}
e^{S(L)}=CL^{\gamma-1}\mu^L\left(1+\frac{a}{L}+\frac{b}{L^{\Delta}}+\dots\right),
\end{equation}
and ignoring the corrections it is 
\begin{equation}
p_{\rm c}(L)=\frac{\mu}{{\cal S}_D C}\left(\frac{L}2\right)^{1-\gamma}.
\label{eq:pc_scaling}
\end{equation}
Here $\gamma$ is another independent universal exponent (cf. Table~\ref{tab:exponent}).
We measured $p_{\rm c}$ for $D=2$ and $D=3$ with FBC. The results are shown in Fig.~\ref{fig:connect_D23} and the fact that $p_{\rm c}/(L/2)^{1-\gamma}$ converges for $L\rightarrow\infty$ demonstrates that the expected scaling from eq. (\ref{eq:pc_scaling}) and in particular the values of $\gamma$ \cite{Clisby_gamma} in Table \ref{tab:exponent} are correct for hard-sphere polymers, too. We also see that the corrections are of the same order as for the end-to-end distance. Interestingly, there seems to be a change in behavior close to $d=0.5$: For both $D=2$ and $D=3$ the 6 lower curves which belong to values $d\in\{0.5,0.6,0.7,0.8,0.9,1.0\}$ are grouped together and are on a logarithmic scale very close to equidistant for all values of $1/L$ or $1/L^{\Delta}$ respectively. For lower values of $d$ we observe greater changes and as expected divergence for $d=0$.

To understand the dependence on the diameter $d$ better, we decided to measure the entropy directly using the newly developed method described above. It samples the state space for many different values of $d$ at the same time and allows to determine its volume by relating it to the known value for $d=0$. Unfortunately, with this algorithm the systems we are able to simulate within reasonable time are smaller than for fixed $d$. Here we present data for $N\le 1024$ and FBC.

%
%
%

\begin{figure}
\begin{center}
\includegraphics[width=0.85\columnwidth]{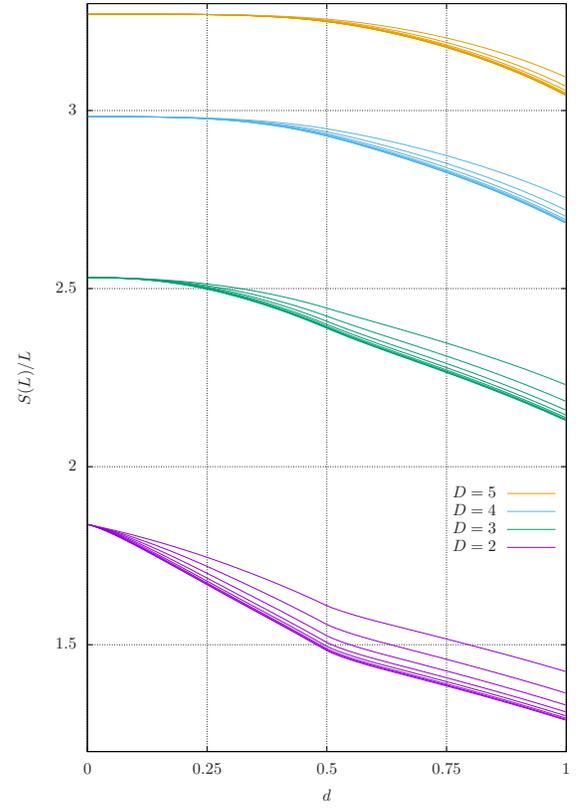}\\
\caption{\small{\label{fig:entropy} Entropy per bond $S/L$ for $D=2,3,4,5$ and\\ $L\in\{7, 15,  31, 63, 127, 255, 511, 1023\}$ with $L$ increasing from top to bottom.}}
\end{center}
\end{figure}

\begin{figure}
\begin{center}
\includegraphics[width=0.99\columnwidth]{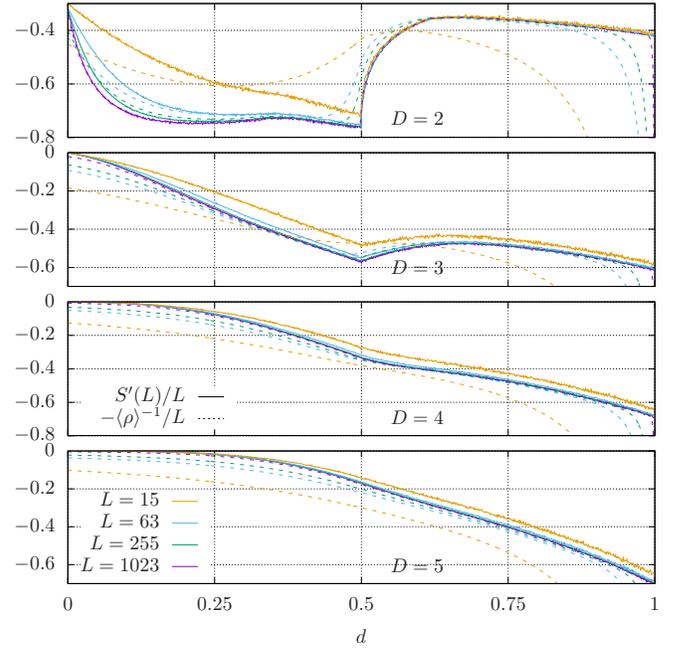}\\
\caption{\small{\label{fig:entropy_d} Numerical derivative of the entropy per bond $S'(d)/L$ and its approximation $-\langle\rho\rangle^{-1}/L$.}}
\end{center}
\end{figure}

\begin{figure}
\begin{center}
\includegraphics[width=0.99\columnwidth]{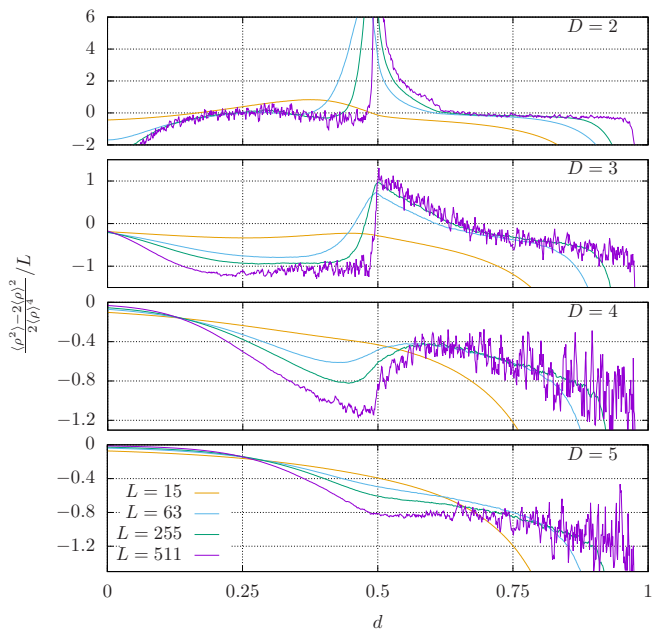}\\
\caption{\small{\label{fig:entropy_dd} Approximation of $S''(d)/L$.}}
\end{center}
\end{figure}

\begin{figure}
\begin{center}
\includegraphics[width=0.99\columnwidth]{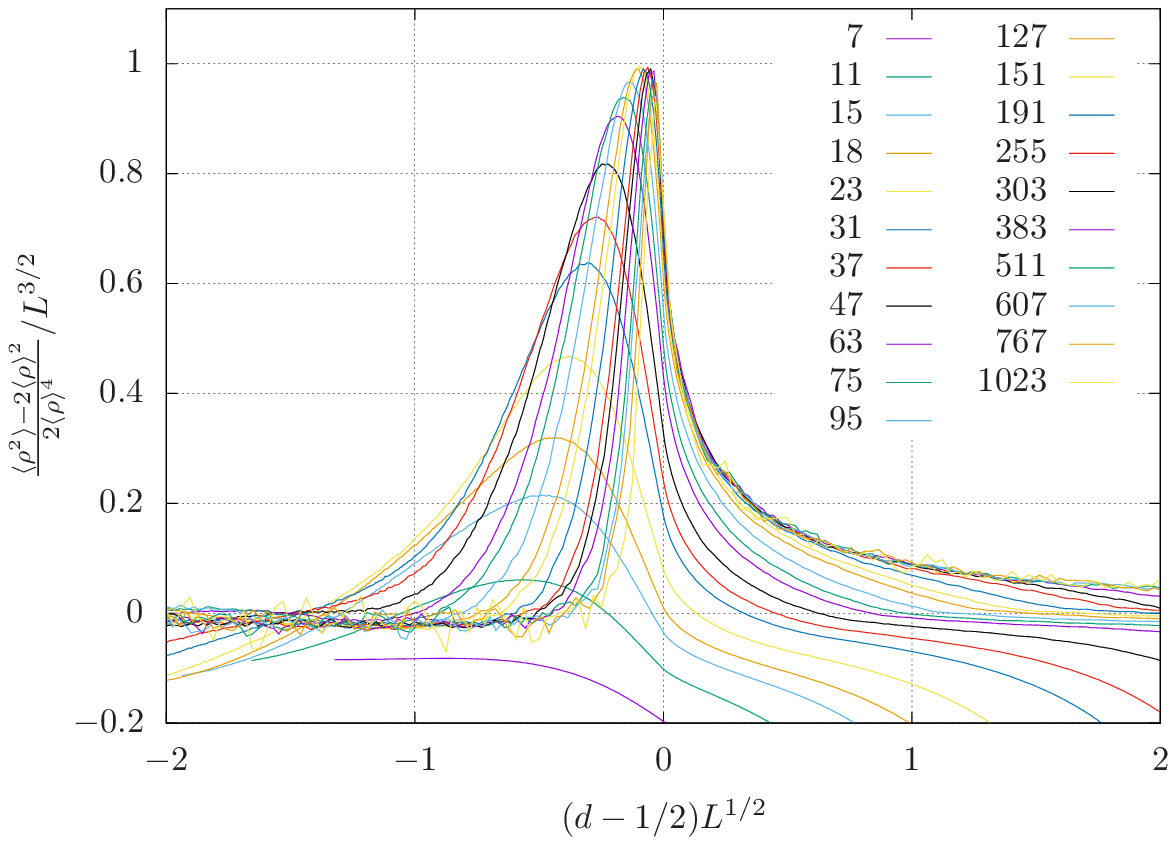}\\
\caption{\small{\label{fig:entropy_D2_dd_peak} Scaling plot of the peak for $D=2$ in Fig.~\ref{fig:entropy_dd}.}}
\end{center}
\end{figure}

In Fig.~\ref{fig:entropy} the entropy per bond $S/L$ is displayed for several chain length and $D=2,3,4,5$. From the MC simulation we only get relative weights and the normalization constant has to be externally provided. Luckily, the values for $d=0$ are known to be $S(d)|_{d=0}=L\ln \cal{S}_D$ with the surface of the $D$-sphere $\cal{S}_D$. We adjust the normalization by vertically shifting the curves such that this condition is fulfilled. We observe that with increasing length $S/L$ decreases slightly, but as expected the curves approach a limiting envelope.

Next we compare in Fig.~\ref{fig:entropy_d} the numerically obtained derivative $S'(d)/L$ to the approximation $-\langle\rho\rangle^{-1}/L$ according to eq.~(\ref{eq:S_deriv_I}). For $L=15$ there are considerable discrepancies which as expected diminish with increasing $L$ until for $L=1023$ the curves are almost identical. We find that the statistical uncertainties are smaller for the approximation. Note that the earlier choice of normalization has no impact here, since it is an added constant to $S/L$.

The approximation of $S''(d)/L$ given in eq.~(\ref{eq:S_deriv_II}) is shown in Fig.~\ref{fig:entropy_dd}. Unfortunately, there is a lot of noise and since we did not get good results for $L=1023$ we only show data for $L\le511$.

For all values of $D$ we observe a signal at $d=1/2$, i.e., for the value that separates the cases where a monomer can be placed in between to bonded ones and those cases where this is no longer possible. For $D=2$ the first derivative has a discontinuity with the associated peak in the second derivative diverging like $L^{3/2}$ (Fig.~\ref{fig:entropy_D2_dd_peak}). For $D=3$ it is the second derivative that becomes discontinuous. The same appears to happen if $D=4$ although the signal is less pronounced and finally for $D=5$ it looks like the third derivative might be discontinuous, however, statistical uncertainties are substantial. There are two other features of interest: if $D=2$ the first derivative of $S$ shows a local maximum at $d\approx0.36$, a value whose significance is not clear to us, and for $D=3$ the second derivative appears to change its slope at $d\approx0.7$, i.e., close to the value of $d=2^{-1/2}$ where the chain ceases to be able to cross itself. We would expect that this value would play a more important role for $D=2$ where it separates values of $d$ for which loops can be formed from those for which this is impossible, but we observe no impact and indeed loops become extremely rare already for diameters smaller than $d=2^{-1/2}$.


The fact that the dominant signals appear to be localized at $d=1/2$ strongly implicates the excluded volume as the direct cause. As illustrated in Fig.~\ref{fig:excl_vol} for $d<1/2$ it growth like $d^D$ while for $d>1/2$ the exponent decreases towards $D-1$. For $D=2$ in particular the boundary of the excluded volume and therefore the rate of its growth actually decreases for $d>1/2$. Of course, looking at an individual configuration only offers limited insight, since with increasing diameter the chain stretches and different configurations dominate the ensemble. The overlap of spheres of excluded volume of beads that are not directly adjacent in the chain will be affected as a consequence.  We have to conclude that the diameter of the beads can only serve as a rough approximation of the strength of the excluded-volume interaction.
This behavior should vanish or at least to be much less pronounced if beads with a softer repulsive potential are used.

Given that the anomalies in the crossover scaling (insets in Fig.~\ref{fig:ree_co_D23}) are maximal in the proximity of $d=1/2$ we find it justified to conclude that they are a secondary effect of the irregularities of $S(d)$ we just discussed.

\section{Conclusion}

\label{sec-Conc}

We investigated hard-sphere polymers in two to five dimensions for the full range of possible sphere diameters.
Thanks to the sophisticated MC methods that we employed it was possible to simulate very long chains with $L\lesssim10^7$.
As expected we found that the scaling behavior known from self-avoiding walks is reproduced in detail by the hard-sphere polymers.
In particular the universal exponent $\nu$ and also the weakly universal ratios of amplitudes of end-to-end distance and radius of gyration for free boundary conditions appear to be the same, thus supporting universality. And while we have not explicitly estimated the exponents $\Delta$ and $\gamma$, using the values established for SAWs during the analysis of our data led to a consistent picture.

We demonstrated that periodic boundary conditions for polymers can be applied within the framework of MC simulations. While universal exponents are unaffected, the coefficients of correction terms are reduced, and new modified values for the (only weakly universal) amplitude ratios are obtained for $D=2$ and $3$. 

It is established knowledge that microscopic details do not matter for the asymptotic scaling behavior of polymers.
This is, of course, the case here as well; for all systems and non-zero sphere diameters we observe the scaling of self-avoiding walks with the appropriate exponents and corrections.
However, at finite chain length the choice of the sphere diameter is relevant.
According to theory for long enough chains the scaling behavior should not depend on length and strength of the repulsive interaction separately, but on a combination of both variables.
We found this prediction to be generally true for $D=2$ and $3$, however, deviations that cannot be explained by finite size occur and are strongest for diameters $d\approx 1/2$.
We suspect that the sphere diameter is an imperfect measure for the strength of the repulsive interaction and as a consequence we lack the correct argument for the scaling function.

For the first time the entropy and its derivatives of hard-sphere polymers were measured to confirm this assumption.
In fact, we find discontinuities localized again at $d=1/2$ for any number of dimensions.
These are clearly a result of the specific geometry of the model and their persistence for increasing length now much more convincingly shows the inadequacy of the microscopic sphere diameter as a descriptor of macroscopic behavior which warrants further research in the future.

\section*{Acknowledgements}

We thank the Deutsche Forschungsgemeinschaft (DFG, German Research Foundation) 
for support through the Sonderforschungsbereich/Transregio under grant No.\ 189\,853\,844 -- SFB/TRR 102 (project B04).


\begin{thebibliography}{99}

\bibitem{deGennes_sausage}
P.-G. de Gennes, J. Phys. Lett. 1985, 46, 639.

\bibitem{Halperin-Goldbart_pearls}
A. Halperin and P.M. Goldbart, Phys. Rev. E 2000, 61, 565.

\bibitem{our_pre2019}
H. Christiansen, S. Majumder, and W. Janke, {\em Phase ordering kinetics of the long-range Ising model\/}, Phys. Rev. E 2019, 99, 011301(R)-1--5.


\bibitem{our_prl2020}
H. Christiansen, S. Majumder, M. Henkel, and W. Janke, {\em Aging in the long-range Ising model\/}, Phys. Rev. Lett. 2020, 125, 180601-1--7.


\bibitem{our_pre2021}
H. Christiansen, S. Majumder, and W. Janke,
{\em Zero-temperature coarsening in the 2D long-range Ising model\/}, Phys. Rev. E 2021, 103, 052122-1--11.


\bibitem{our_epjb2020a}
S. Majumder, H. Christiansen, and W. Janke,{\em Understanding nonequilibrium scaling laws governing collapse of a polymer\/}, Invited Colloquium, Eur. Phys. J. B 2020, 93, 142-1--19.


\bibitem{our_macromolecules2019}
S. Majumder, U.H.E. Hansmann, and W. Janke,{\em Pearl-necklace-like local ordering drives polypeptide collapse\/}, Macromolecules 2019, 52, 5491--5498.


\bibitem{our_jpcs2019a} S. Majumder, H. Christiansen, and W. Janke,{\em Dissipative dynamics of a single polymer in solution: A Lowe-Andersen approach\/}, J. Phys.: Conf. Ser. 2019, 1163, 012072-1--6.


\bibitem{our_viscous-tobe1}
S. Majumder, H. Christiansen, and W. Janke, in preparation.

\bibitem{our_smat2021}
S. Paul, S. Majumder, and W. Janke, {\em Motion of a polymer globule with Vicsek-like activity: From super-diffusive to ballistic behavior\/}, Soft Materials 2021, 19, 306--315.


\bibitem{our_sm2022a}
S. Paul, S. Majumder, S.K. Das, and W. Janke,
{\em Effects of alignment activity on the collapse kinetics of a flexible polymer\/}, Soft Matter 2022, 18, 1978--1990.

\bibitem{our_sm2022b}
S. Paul, S. Majumder, and W. Janke, {\em Activity mediated globule to coil transition of a flexible polymer in poorsolvent\/}, Soft Matter 2022, 18, 6392--6403.


\bibitem{our_jpcs2022a} S. Paul, S. Majumder, and W. Janke, {\em Role of temperature and alignment activity on kinetics of coil-globule transition of a flexible polymer\/}, J. Phys.: Conf. Ser. 2022, 2207, 012027-1--6.


\bibitem{ours_macromolecules2021}
S. Majumder, M. Marenz, S. Paul, and W. Janke, {\em Knots are generic stable phases in semiflexible polymers\/}, Macromolecules 2021, 54, 5321--5334.


\bibitem{pa_iba}
Y. Iba, Trans. Jpn. Soc. Artif. Intell. 2001, 16, 279.

\bibitem{pa_hukushima-iba}
K. Hukushima and Y. Iba, in {\em The Monte Carlo Method in the Physical Sciences: Celebrating the 50th Anniversary of the Metropolis Algorithm\/}, edited by J.E. Gubernatis, AIP Conf.\
Proc.\ No.\ 690 (AIP, New York, 2003), pp.\ 200--206.

\bibitem{pa_machta}
J. Machta, Phys. Rev. E 2010, 82, 026704.

\bibitem{martin2021}
M. Weigel, L.Yu. Barash, L.N. Shchur, and W. Janke, {\em Understanding population annealing Monte Carlo simulations\/}, Phys. Rev. E 2021, 103, 053301-1--24.


\bibitem{our_prl2019}
H. Christiansen, M. Weigel, and W. Janke,
{\em Accelerating molecular dynamics simulations with population annealing\/}, Phys. Rev. Lett. 2019, 122, 060602-1--5.


\bibitem{our_jpcs2019b}
H. Christiansen, M. Weigel, and W. Janke, {\em Population annealing molecular dynamics with adaptive temperature steps\/}, J. Phys.: Conf. Ser. 2019, 1163, 012074-1--6.


\bibitem{our_jpcs2022b}
H. Christiansen, M. Weigel, and W. Janke, {\em Simulating met-enkephalin with population annealing molecular dynamics\/}, J. Phys.: Conf. Ser. 2022, 2241, 012006-1--7.


\bibitem{lev2017}
L.Yu. Barash, M. Weigel, M. Borovsk{\'y}, W. Janke, and L.N. Shchur, {\em GPU accelerated population annealing algorithm\/}, Comput. Phys. Commun. 2017, 220, 341--350.


\bibitem{lev2019}
L. Shchur, L. Barash, M. Weigel, and W. Janke, {\em Population annealing and large scale simulations in statistical mechanics\/}, Communications in Computer and Information Science (CCIS) 2019, 965, 354--366.


\bibitem{our_proteinG-tobe}
H. Christiansen, F. M\"uller, D. Gessert, M. Weigel, and W. Janke, {\em Massively parallel simulation of peptides with explicit solvent\/}, in preparation.

\bibitem{our_pre2020}
F. M\"uller, S. Schnabel, and W. Janke, {\em Nonflat histogram techniques for spin glasses\/}, Phys. Rev. E 2020, 102, 053303-1--8.


\bibitem{muca1}
B.A. Berg and T. Neuhaus, 
Phys. Lett. B 1991, 267, 249.

\bibitem{muca2}
B.A. Berg and T. Neuhaus, 
Phys. Rev. Lett. 1992, 68, 9.

\bibitem{flat-review_wj-paul}
W. Janke and W. Paul, {\em Thermodynamics and structure of macromolecules from flat-histogram Monte Carlo simulations\/} (invited review), Soft Matter 2016, 12, 642.

\bibitem{wl}
F. Wang and D. P. Landau, 
Phys. Rev.  Lett. 2001, 86, 2050.

\bibitem{our_cpc2021}
S. Schnabel and W. Janke, {\em Wang-Landau simulations with non-flat distributions\/}, Comput. Phys. Commun. 2021, 267, 108071-1--5.


\bibitem{our_epjb2020b}
S. Schnabel and W. Janke,
Counting metastable states of Ising spin glasses on hypercubic lattices,
Eur. Phys. J. B 2020, 93, 53-1--7.


\bibitem{our_HP-tobe}
S. Schnabel and W. Janke,
in preparation.

\bibitem{our_cpc2020}
S. Schnabel and W. Janke,
Polymer simulation by means of tree data-structures and a parsimonious 
Metropolis algorithm,
Comput. Phys. Commun. 2020, 256, 107414-1--9.


\bibitem{our_arxiv2022}
F. M\"uller, H. Christiansen, S. Schnabel, and W. Janke,
{\em Fast, hierarchical, and adaptive algorithm for Metropolis Monte Carlo simulations of long-range interacting systems\/},
arXiv:2207.14670 (physics.comp-ph, cond-mat.stat-mech), 2022. 


\bibitem{our_prl2022}
F. M\"uller, H. Christiansen, and W. Janke,
{\em Phase-separation kinetics in the two-dimensional long-range Ising model\/}, Phys. Rev. Lett. 2022, 129, 240601-1--6.

\bibitem{our_jpcp2022}
S. Schnabel and W. Janke,
{\em Fast simulation of a large polymer with untruncated interaction near the 
collapse transition\/},
J. Phys.: Conf. Ser. 2022, 2241, 012005-1--8.


\bibitem{TSAW}
N. Madras and G. Slade, {\em The Self-Avoiding Walk\/}, Birkh\"auser, Boston 1993.

\bibitem{Sokal_book}
R. Fern\'andez, J. Fr\"ohlich, and A. D. Sokal, {\em Random Walks, Critical Phenomena, and Triviality in Quantum Field Theory\/}, Springer, Berlin 1992.

\bibitem{Clisby2}
N. Clisby, J. Stat. Phys. 2010, 140, 349.

\bibitem{Clisby3}
N. Clisby and B. D\"unweg, Phys. Rev. E 2016, 94, 052102.

\bibitem{Deng_saw_4d}
S. Fang, Y. Deng, and Z. Zhou, Phys. Rev. E 2021, 104, 064108.

\bibitem{Kremer}
K. Kremer, A. Baumg\"artner, and K. Binder, Z. Physik B 1981, 40, 331.

\bibitem{Sokal}
N. Madras and A. D. Sokal, J. Stat. Phys. 1988, 50, 109.

\bibitem{EndlessSAW}
N. Clisby, J. Phys. A 2013, 46, 235001.

\bibitem{Clisby1}
N. Clisby, Phys. Rev. Lett. 2010, 104, 055702.


\bibitem{saw_enum_D3}
R. D. Schram, G. T. Barkema, and R. H. Bisseling, J. Stat. Mech. 2011, P06019.

\bibitem{saw_enum_D2_B}
I. Jensen, arXiv:1309.6709 (math-ph), 2013.

\bibitem{saw_enum_D2}
I. Jensen, J. Phys. A 2004, 37, 5503.

\bibitem{Clisby_gamma}
N. Clisby, J. Phys. A 2017, 50, 264003.

\bibitem{JointCut}
S. Caracciolo, A. Pelissetto, and A. D. Sokal, J. Stat. Phys. 1992, 67, 65.

\bibitem{Kleinert}
H. Kleinert and V. Schulte-Frohlinde, Critical Properties of $\phi^4$-Theories, World Scientific Publishing, Singapore 2001.

\bibitem{Shalaby}
A. M. Shalaby, Eur. Phys. J. C 2021, 81, 87.

\bibitem{Duplantier}
B. Duplantier, Nucl. Phys. B 1986, 275, 319.

\bibitem{saw_4D_Grassberger}
P. Grassberger, R. Hegger, and L. Sch\"afer, J. Phys. A 1994, 27, 7265.

\bibitem{saw_D4_Clisby}
N. Clisby, J. Stat. Phys. 2018, 172, 477.

\bibitem{RKenna}
B. Berche, T. Ellis, Y Holovatch, and R. Kenna, SciPost Phys. Lect. Notes 2022, 60.

\bibitem{saw_D5_Prellberg}
A. L. Owczarek and T. Prellberg, J. Phys. A  2001, 34, 5773.

\bibitem{Li_Madras_Sokal}
B. Li, N. Madras, and  A. D. Sokal, J. Stat. Phys. 1995, 80, 661.


%


\end{thebibliography}
\end{document}